\documentclass[submission,copyright,creativecommons,sharealike,noncommercial]{eptcs}
\providecommand{\event}{QPL 2017}
\pdfoutput=1

\usepackage{mathtools} % Loads and extends amsmath
\usepackage{amssymb} % Extram mathematical symbols (loads amsfonts)
\usepackage{bbold} % Sans-serif blackboard bold font for numbers
\usepackage{amsthm} % Theorem environments
\usepackage{stmaryrd} % Some maths symbols for logic and computer science
\usepackage{relsize} % Additional relative sizes for fonts
\usepackage{microtype} % Improves appearance of writing
\usepackage{csquotes} % Environments for quotes
\usepackage{cancel}

\usepackage[nocompress]{cite} % Comment out if using natbib or apacite
\usepackage{graphicx} % Import of graphics
\usepackage[usenames,dvipsnames]{xcolor} % Introduces colour names
\usepackage{tikz} % TikZ
\usepackage{circuitikz} % TikZ circuit diagrams 
\usetikzlibrary{
	arrows,
	shapes,
	decorations,
	intersections,
	backgrounds,
	positioning,
	circuits.ee.IEC
	}

		\newcounter{theorem_c} % Unified coutner for all theorem environments...
		\numberwithin{theorem_c}{section} % ... numbered within sections
		\numberwithin{equation}{section} % Equations are also numbered within sections (but have a separate counter)
		\theoremstyle{plain} 
		\newtheorem{theorem}[theorem_c]{Theorem}
		
		\newtheorem{lemma}[theorem_c]{Lemma}
		\newtheorem{corollary}[theorem_c]{Corollary}
		
		\newtheoremstyle{exampstyle}
		  {2mm} % Space above
		  {2mm} % Space below
		  {\itshape} % Body font
		  {} % Indent amount
		  {\bfseries} % Theorem head font
		  {.} % Punctuation after theorem head
		  {.5em} % Space after theorem head
		  {} % Theorem head spec (can be left empty, meaning `normal')

		\theoremstyle{exampstyle}
		\newtheorem{definition}[theorem_c]{Definition}
		
		\newtheorem{remark}[theorem_c]{Remark}

	\newcommand{\inlineQuote}[1]{\textquotedblleft #1\textquotedblright} % Left-right quotes surrounding #1
	\newcommand{\goodchi}{\protect\raisebox{2pt}{$\chi$}} % A chi letter raised at line level, for those instances where alignment is an issue
	\newcommand{\goodvdots}{\protect\raisebox{7pt}{\vdots}} % A \vdots simbol raised, for use in tikz nodes where vertical alignment is a problem 

	\newcommand{\naturals}{\mathbb{N}} % Set of natural numbers
	\newcommand{\integers}{\mathbb{Z}} % Set of interer numbers
	\newcommand{\rationals}{\mathbb{Q}} % Set of rational numbers
	\newcommand{\reals}{\mathbb{R}} % Set of real numbers
	\newcommand{\complexs}{\mathbb{C}} % Set of complex numbers
	\newcommand{\integersMod}[1]{\mathbb{Z}_{#1}} % Set/group/ring of integers mod #1
	\newcommand{\splitComplexs}{\mathbb{C}[\sqrt{1}]} % Ring of split complex numbers
	\newcommand{\finiteField}[1]{\mathbb{F}_{\!\!#1}} % Finite field
	\newcommand{\singletonSet}{\textbf{1}}
	
	\newcommand{\ord}[1]{{\operatorname{ord} #1}}
	\newcommand{\sgn}[2]{{\operatorname{sgn}_{#1} #2}}

	\newcommand{\suchthat}[2]{\left\{#1 \: \middle\vert \: #2\right\}} % Set of elements #1 such that condition #2 holds 
		\newcommand{\ket}[1]{\vert #1 \rangle} % Ket labelled #1
		\newcommand{\bra}[1]{\langle #1 \vert} % Bra labelled #1
		\newcommand{\decohSym}{\operatorname{dec}} % Decoherence map symbol
		\newcommand{\decoh}[1]{\decohSym_{#1}} % Decoherence map (for a given observable)
		\newcommand{\CPMdoubled}[1]{\textbf{double}\left[#1\right]}

		\newcommand{\SpaceH}{\mathcal{H}} 
		\newcommand{\SpaceG}{\mathcal{G}}
		\newcommand{\SpaceK}{\mathcal{K}}

		\newcommand{\isom}{\cong} % Isomorphism
		\newcommand{\id}[1]{id_{#1}} % Identity morphism of object #1
		\newcommand{\CMonCategory}{\operatorname{CMon}} % Category of commutative monoids
		\newcommand{\RMatCategory}[1]{#1\operatorname{-Mat}} % Category of finite sets and matrices valued in a semiring R 
		\newcommand{\fHilbCategory}{\operatorname{fHilb}} % Category of finite-dimensional Hilbert spaces
		\newcommand{\fdHilbCategory}{\fHilbCategory} % Alias for the category of finite-dimensional Hilbert spaces (for retrocompatibility)
		\newcommand{\fSetCategory}{\operatorname{fSet}} % Category of finite sets and total functions
		\newcommand{\CategoryC}{\mathcal{C}}

		\newcommand{\obj}[1]{\operatorname{obj} \, #1} % Set of objects of category #1
		\newcommand{\CPMCategory}[1]{\operatorname{CPM}[#1]} % Selinger's CPM category
		\newcommand{\CPStarCategory}[1]{\operatorname{CP}^\ast[#1]} % Chris's CP star construction 
		\newcommand{\classicalSubcategory}[1]{#1_{K}} % The classical subcategory in a probabilistic theory 

	\newcommand{\Xcolour}{Red}
	\newcommand{\Zcolour}{YellowGreen}
	\newcommand{\Xaltcolour}{Purple}
	
	\newcommand{\Zaltcolour}{Cyan}
	
	\newcommand{\Dcolour}{black!80}
	\newcommand{\Xbwcolour}{black!80}
	\newcommand{\Zbwcolour}{white}
	\newcommand{\Ybwcolour}{black!15}
	\newcommand{\Wbwcolour}{black!50}
	\newcommand{\trace}[1]{\hbox{\begin{tikzpicture} [scale=1.2,transform shape, rotate = -90] %% DO NOT CHANGE

\def\deltax{0.3} %% CAN BE CHANGED
\def\deltay{0.5} %% DO NOT CHANGE

\path[use as bounding box] (-\deltax,-0.7*\deltay) rectangle (\deltax,0.3*\deltay);

\node (mult) at (0,0.3*\deltay) [upground,scale=0.5] {};
\node (mult_label_in) at (0,-0.7*\deltay) {};
\draw[-] (mult_label_in) to (mult);

\end{tikzpicture}}\!_{#1}} % Trace
	\tikzset{
	  rectangle with rounded corners north west/.initial=4pt,
	  rectangle with rounded corners south west/.initial=4pt,
	  rectangle with rounded corners north east/.initial=4pt,
	  rectangle with rounded corners south east/.initial=4pt,
	}
	\makeatletter
	\pgfdeclareshape{rectangle with rounded corners}{
	  \inheritsavedanchors[from=rectangle] % this is nearly a rectangle
	  \inheritanchorborder[from=rectangle]
	  \inheritanchor[from=rectangle]{center}
	  \inheritanchor[from=rectangle]{north}
	  \inheritanchor[from=rectangle]{south}
	  \inheritanchor[from=rectangle]{west}
	  \inheritanchor[from=rectangle]{east}
	  \inheritanchor[from=rectangle]{north east}
	  \inheritanchor[from=rectangle]{south east}
	  \inheritanchor[from=rectangle]{north west}
	  \inheritanchor[from=rectangle]{south west}
	  \backgroundpath{% this is new
	    \southwest \pgf@xa=\pgf@x \pgf@ya=\pgf@y
	    \northeast \pgf@xb=\pgf@x \pgf@yb=\pgf@y
	    \pgfkeysgetvalue{/tikz/rectangle with rounded corners north west}{\pgf@rectc}
	    \pgfsetcornersarced{\pgfpoint{\pgf@rectc}{\pgf@rectc}}
	    \pgfpathmoveto{\pgfpoint{\pgf@xa}{\pgf@ya}}
	    \pgfpathlineto{\pgfpoint{\pgf@xa}{\pgf@yb}}
	    \pgfkeysgetvalue{/tikz/rectangle with rounded corners north east}{\pgf@rectc}
	    \pgfsetcornersarced{\pgfpoint{\pgf@rectc}{\pgf@rectc}}
	    \pgfpathlineto{\pgfpoint{\pgf@xb}{\pgf@yb}}
	    \pgfkeysgetvalue{/tikz/rectangle with rounded corners south east}{\pgf@rectc}
	    \pgfsetcornersarced{\pgfpoint{\pgf@rectc}{\pgf@rectc}}
	    \pgfpathlineto{\pgfpoint{\pgf@xb}{\pgf@ya}}
	    \pgfkeysgetvalue{/tikz/rectangle with rounded corners south west}{\pgf@rectc}
	    \pgfsetcornersarced{\pgfpoint{\pgf@rectc}{\pgf@rectc}}
	    \pgfpathclose
	 }
	}
	\makeatother

	\tikzset{->-/.style={decoration={markings,mark=at position #1 with {\arrow{>}}},postaction={decorate}}}
	\tikzset{-<-/.style={decoration={markings,mark=at position #1 with {\arrow{<}}},postaction={decorate}}}

	\tikzstyle{every picture}=[baseline=-0.25em,scale=0.5]
	\pgfdeclarelayer{edgelayer}
	\pgfdeclarelayer{nodelayer}
	\pgfsetlayers{edgelayer,nodelayer,main}

	\tikzstyle{box} = [draw,shape=rectangle,inner sep=2pt,minimum height=6mm,minimum width=6mm,fill=white] 
	\tikzstyle{boxlarge} = [draw,shape=rectangle,inner sep=2pt,minimum height=1.5cm,minimum width=8mm,fill=white] 
	\tikzstyle{boxLarge} = [draw,shape=rectangle,inner sep=2pt,minimum height=2cm,minimum width=10mm,fill=white] 
	\tikzstyle{boxsmall} = [draw,shape=rectangle,inner sep=2pt,minimum height=3mm,minimum width=3mm,fill=white] % small in all directions. Might one day use boxnarrow for small in the largeness direction only.
	\tikzstyle{dot} = [inner sep=0mm,minimum width=3mm,minimum height=3mm,draw,shape=circle,text depth=-0.1mm]
	\tikzstyle{Zbwdot} = [dot, fill=\Zbwcolour]
	\tikzstyle{Xbwdot} = [dot, fill=\Xbwcolour]
	\tikzstyle{Ybwdot} = [dot, fill=\Ybwcolour]
	\tikzstyle{Wbwdot} = [dot, fill=\Wbwcolour]
	\tikzstyle{antipode} = [boxsmall] 

	\tikzstyle{state} = [draw, rectangle with rounded corners,
	  rectangle with rounded corners north west=8pt,
	  rectangle with rounded corners south west=8pt,
	  rectangle with rounded corners north east=0pt,
	  rectangle with rounded corners south east=0pt,
	,inner sep=2pt,minimum height=6mm,minimum width=6mm,fill=white]
	\tikzstyle{statelarge} = [draw, rectangle with rounded corners,
	  rectangle with rounded corners north west=8pt,
	  rectangle with rounded corners south west=8pt,
	  rectangle with rounded corners north east=0pt,
	  rectangle with rounded corners south east=0pt,
	,inner sep=2pt,minimum height=1.5cm,minimum width=8mm,fill=white]
	\tikzstyle{stateLarge} = [draw, rectangle with rounded corners,
	  rectangle with rounded corners north west=8pt,
	  rectangle with rounded corners south west=8pt,
	  rectangle with rounded corners north east=0pt,
	  rectangle with rounded corners south east=0pt,
	,inner sep=2pt,minimum height=2cm,minimum width=8mm,fill=white]
	\tikzstyle{effect} = [draw, rectangle with rounded corners,
	  rectangle with rounded corners north west=0pt,
	  rectangle with rounded corners south west=0pt,
	  rectangle with rounded corners north east=8pt,
	  rectangle with rounded corners south east=8pt,
	,inner sep=2pt,minimum height=6mm,minimum width=6mm,fill=white]
	\tikzstyle{scalar}=[diamond,draw,inner sep=1pt,font=\small,fill=white]

	\tikzstyle{cdnode}=[fill=white]
	\tikzstyle{labelnode}=[fill=white]
	\tikzstyle{tightlabelnode}=[fill=white,inner sep = 0.1mm]
	\tikzstyle{none}=[inner sep=0pt]
	\tikzstyle{whiteline}=[-, line width=4pt, draw=white]
	\tikzstyle{trace}=[circuit ee IEC,thick,ground,scale=2.5]
	\tikzstyle{cotrace}=[circuit ee IEC,thick,ground,rotate=180,scale=2.5]
	\tikzstyle{upground}=[circuit ee IEC,thick,ground,rotate=90,scale=2.5]
	\tikzstyle{downground}=[circuit ee IEC,thick,ground,rotate=-90,scale=2.5]

	\tikzstyle{doubled} = [line width=1.8pt] % [line width=1.6pt] % [very thick]
	
	\tikzstyle{empty diagram}=[draw=gray!40!white,dashed,shape=rectangle,minimum width=1cm,minimum height=1cm]

\title{Fantastic Quantum Theories and Where to Find Them}
\author{
	Stefano Gogioso\\
	University of Oxford \\
	\texttt{stefano.gogioso@cs.ox.ac.uk}
}

\def\titlerunning{Fantastic Quantum Theories and Where to Find Them}
\def\authorrunning{S. Gogioso}

\begin{document}

\maketitle

\begin{abstract}
	We present a uniform framework for the treatment of a large class of toy models of quantum theory. Specifically, we will be interested in theories of wavefunctions valued in commutative involutive semirings, and which give rise to some semiring-based notion of classical non-determinism via the Born rule. The models obtained with our construction possess many of the familiar structures used in Categorical Quantum Mechanics. We also provide a bestiary of increasingly exotic examples: some well known, such as real quantum theory and relational quantum theory; some less known, such as hyperbolic quantum theory, p-adic quantum theory and ``parity quantum theory''; and some entirely new, such as ``finite-field quantum theory'' and ``tropical quantum theory''. As a further bonus, the measurement scenarios arising within these theories can be studied using the sheaf-theoretic framework for non-locality and contextuality. Their computational complexity can similarly be studied within existing frameworks for affine and unitary circuits over commutative semirings.
\end{abstract}
\vspace{-0.3cm}

\section{Introduction} 
\label{section_introduction}

The construction of toy models plays a key role in many foundational efforts across mathematics, physics and computer science. In the foundations of quantum theory, toy models help to understand which abstract structural features of quantum systems---and their interface to the classical world---are involved in providing different kinds of non-classical behaviour. In turn, this informs practical research into quantum computation and communication technology, helping to cut down the noise and focus on those features that truly contribute to quantum advantage.

The categorical and diagrammatic methods from Categorical Quantum Mechanics (CQM) \cite{Abramsky2004,Coecke2011,Coecke2015,Heunen2016a,Coecke2016a} have proven particularly well suited to the construction and study of toy models, with the majority of efforts focussed on Spekkens's toy model \cite{Spekkens2007,Coecke2012a,Backens2015} and the more general \textit{relational quantum theory}\footnote{Not to be confused with the \textit{relational quantum mechanics} of Rovelli \cite{Rovelli1996}.} \cite{Pavlovic2009,Abramsky2012,Evans2009,Coecke2012c,Gogioso2015g,Zeng2015,Marsden,Heunen2015}. In the same years, models have been developed within a variety of other frameworks: examples include \textit{real quantum theory}, of special interest in the context of generalised/operational probabilistic theories and the study of Jordan algebras \cite{Jordan1934,Araki1980,Wootters1990,Chiribella2010,Baez2012,Belenchia2012,Wilce2016}, \textit{hyperbolic quantum theory} \cite{Khrennikov2003,Khrennikov2010,Nyman2011}, \textit{$p$-adic quantum theory} \cite{Vladimirov1989,Ruelle1989,Khrennikov1991,Khrennikov1993,Palmer2016,Palmer2016a}, and \textit{modal quantum theory} \cite{Schumacher2012,Schumacher2016,DeBeaudrap2014}.

When constructing a toy model, it is essential to consider both the quantum side and the corresponding quantum-classical interface: indeed, many toy models result in notions of classical non-determinism which are different from the conventional probabilistic one, and special care needs to be taken in order to achieve a consistent treatment of classical systems. Examples of this phenomenon include the possibilistic non-determinism arising from Spekkens's toy model and relational quantum theory \cite{Coecke2012a,Abramsky2012,Abramsky2011,Abramsky2013,Abramsky2012c}, the $p$-adic non-determinism arising from $p$-adic quantum theory \cite{Khrennikov1993}, and the signed probabilities arising from hyperbolic quantum theory \cite{Abramsky2011,Abramsky2014}. 

Because of this issue, we adopt a recently developed framework, that of Categorical Probabilistic Theories \cite{Gogioso2017a}, which can simultaneously treat quantum-like systems and classical systems endowed with generic semiring-based notion of non-determinism. Categorical Probabilistic Theories have been introduced with the intent of bridging the gap between CQM and Operational Probabilistic Theories (OPTs) \cite{Chiribella2010,Hardy2009,Chiribella2011,Chiribella2014,Hardy2016}: they aim to provide categorical and diagrammatic methods in the style of CQM to talk about the problems that OPTs are concerned with. As a side-product of their abstract categorical formulation, these theories natively admit a general, semiring-based notion of classical non-determinism, and are therefore perfect to construct and study exotic toy models of quantum theory\footnote{Anything which is not probabilistic should be deemed to be more or less exotic, in the sense that conventional wisdom about classical systems might fail in one way or another. This includes all toy models mentioned above, except for real quantum theory.}.  

In this work, we focus our attention on a very large class of finite-dimensional quantum-like theories, where wavefunctions of complex amplitude are replaced with wavefunctions valued in some arbitrary commutative semiring $S$ with involution. In the quantum-classical transition, probabilities are still deemed to arise via the Born rule, and as a consequence classical non-determinism is naturally and necessarily modelled by the semiring $R$ of \textit{positive elements} for $S$, generalising the traditional probabilistic semiring $\reals^+$ (which is the semiring of positive elements for $\complexs$, with complex conjugation as involution).

As an underlying model for $S$-valued wavefunctions we consider the category $\RMatCategory{S}$, with objects in the form $S^X$ for all finite sets $X$, and morphisms $S^X \rightarrow S^Y$ given by $S^{Y \times X}$, the $S$-semimodule of $Y$-by-$X$ matrices with values in $S$ (equipped with matrix composition and identities). The category $\RMatCategory{S}$ generalises $\fdHilbCategory \simeq \RMatCategory{\complexs}$, and is a dagger compact category with Kronecker product as tensor product and conjugate\footnote{With respect to the involution $^\ast$ of the commutative semiring $S$.} transpose as dagger. We model mixed-state quantum theory using the CPM construction \cite{Selinger2007}, and the full quantum-classical theory using the dagger-idempotent variant of the CP* construction presented in Ref. \cite{Selinger2008} (instead of the one from Refs. \cite{Coecke2012d,Coecke2014a}), which results in quantum-classical transitions governed by a categorical flavour of the Born rule. Hence our full theory will be described by $\CPStarCategory{\RMatCategory{S}}$, which we will show to be an $R$-probabilistic theory in the sense of Ref. \cite{Gogioso2017a}, and hence amenable for treatment in the framework of Categorical Probabilistic Theories. Furthermore, this formulation is compatible with the framework developed in Ref. \cite{DeBeaudrap2014}, which can therefore be used to investigate natural notions of computational complexity for the toy models we present.\vspace{-0.5cm}

\paragraph{Synopsis.} We begin in Sections \ref{section_probabilisticTheories} and \ref{section_quantumLikeTheories} by summarising the structure of Categorical Probabilistic Theories, and by introducing our general framework for the construction of theories of wavefunctions valued in commutative involutive semirings. The remainder of the work is then dedicated to showing that our new framework is expressive enough to cover all the interesting toy models which we mentioned. 

In Sections \ref{section_realQT} and \ref{section_relationalQT} we briefly cover real quantum theory and relational quantum theory, two models already dear to the CQM and OPT communities. In Sections \ref{section_hyperbolicQT}, \ref{section_parityQT} and \ref{section_pAdicQT} we cover hyperbolic quantum theory, parity quantum theory and $p$-adic quantum theory, already known in some specialised circles but not yet part of the categorical bestiary. Hyperbolic quantum theory is particularly interesting: it is a quasi-probabilistic theory which can support at least as many protocols as real quantum theory (e.g. Simon's problem can be solved efficiently), but at the same time it is entirely local thanks to negative probabilities. Furthermore, parity quantum theory was already known as the $\integersMod{2}$ case of modal quantum theory, although a different interpretation is given here to its classical non-determinism. 

In Sections \ref{section_finitefieldQT} and \ref{section_tropicalQT} we present two new families of toy models: finite-field quantum theory and tropical quantum theory, based on finite fields and tropical semirings respectively. Finite-field quantum theory proves to be especially interesting: it possesses a rich phase group, and can support a number of non-trivial quantum protocols and arguments, but at the same time it is entirely local. Furthermore, its pure quantum states are points of finite projective spaces, and can be studies using tools from finite geometry. Tropical quantum theory, on the other hand, turns out to be an extension of boolean quantum theory alternative to relational quantum theory, with which it shares many structural and operational traits.

\newpage
\section{Categorical Probabilistic Theories}
\label{section_probabilisticTheories}

The main intuition behind a generalised, semiring-based notion of classical non-deterministic systems is borrowed from computer science, where the use of (commutative) semirings to model resources used by automata is commonplace. We look at probabilities in physics as a resource modelling non-determinism of classical systems, with properties captured by those of the commutative semiring $\reals^+$: from this perspective, it makes sense to study what classical non-determinism looks like when $\reals^+$ is replaced by some other commutative semiring $R$. Interesting alternative choices for $R$ which already appeared in the literature include the boolean semiring $\mathbb{B}$ and other locales (in relational quantum theory), the quasi-probabilistic semiring $\reals$ (a field, in hyperbolic quantum theory), the $p$-adic semiring $Q_p$ (another field, in $p$-adic quantum theory), and finite fields (in modal quantum theory). We will refer to classical systems with non-determinism governed by a commutative semiring $R$ as \textbf{classical $R$-probabilistic systems}. 

One of the reason for the wide adoption of semirings in mathematics is that they capture the bare minimum algebraic structure required by matrix multiplication, with commutativity being a necessary addendum when a symmetric tensor product of matrices is of interest (as is the case in many physical applications). As our \textbf{category of classical $R$-probabilistic systems} we take the category $\RMatCategory{R}$ of free finite-dimensional $R$-semimodules and $R$-linear maps between them: objects are in the form $R^X$ with $X$ finite sets, and morphisms $R^X \rightarrow R^Y$ form the free finite-dimensional $R$-semimodule of $Y$-by-$X$ matrices with values in $R$. The category $\RMatCategory{R}$ is a compact closed symmetric monoidal category, with Kronecker product of matrices as tensor product. It is enriched in itself, and hence in commutative monoids ($\CMonCategory$-enriched), so that each homset $R^X \rightarrow R^Y$ comes with a \textbf{mixing operation} $+$ and an \textbf{impossible process} $0$. The category $\RMatCategory{R}$ contains the category $\fSetCategory$ of finite sets and functions (the category of \textbf{classical deterministic systems}) as a subcategory, and from $\fSetCategory$ it inherits an environment structure $(\trace{\,R^X}: R^X \rightarrow R^\singletonSet)_{X}$ \cite{Coecke2008a,Coecke2014a} given by the \textbf{discarding maps} $\trace{\,R^X} := (p_x)_x \mapsto \sum_x p_x$.

In Ref. \cite{Gogioso2017a}, it is argued that the minimal requirements for a categorical probabilistic theory should include: (i) the explicit existence of classical systems\footnote{So that the interface between classical and non-classical systems can be talked about in a compositional way. This includes, for example, classical control, classical outcomes, preparations and measurements.}; (ii) the extendibility of probabilistic mixing to all systems\footnote{So that, for example, classical probabilistic control and marginalisation over classical outcomes are possible for all processes.}; (iii) the possibility of defining a meaningful notion of local state and discarding of systems\footnote{Which are absolutely fundamental in most applications to quantum foundations and quantum protocols (but I acknowledge that Everettians and other faithful of the Church of the Larger Hilbert Space might disagree with me on this point).}.

\begin{definition}[\textbf{$R$-probabilistic Theory}]\hfill\\
An \textbf{$R$-probabilistic theory} is a symmetric monoidal category $\CategoryC$ which satisfies the following requirements.
\begin{enumerate}
	\item[(i)] There is a full sub-SMC of $\CategoryC$, denoted by $\classicalSubcategory{\CategoryC}$, which is equivalent to $\RMatCategory{R}$. 
	\item[(ii)] The SMC $\CategoryC$ is enriched in commutative monoids, and the enrichment on the subcategory $\classicalSubcategory{\CategoryC}$ coincides with the one given by the linear structure of $\RMatCategory{R}$.
	\item[(iii)] The SMC $\CategoryC$ comes with an environment structure, i.e. with a family $(\trace{\SpaceH}:\SpaceH \rightarrow \singletonSet)_{\SpaceH \in \obj{\CategoryC}}$ of morphisms which satisfy the following requirements:
		\begin{equation}\label{environmentStructure}
		\begin{tikzpicture}[scale=0.8]
	\begin{pgfonlayer}{nodelayer}
		\node [style=labelnode] (0) at (-7.5, 0) {$=$};
		\node [style=labelnode, inner sep=0.1 mm] (1) at (-13.5, 0) {$\SpaceH \otimes \SpaceG$};
		\node [style=trace] (2) at (-9.5, 0) {};
		\node [style=labelnode, inner sep=0.1 mm] (3) at (-5.5, 0.75) {$\SpaceH$};
		\node [style=trace] (4) at (-2.5, 0.75) {};
		\node [style=labelnode, inner sep=0.1 mm] (5) at (-5.5, -0.75) {$\SpaceG$};
		\node [style=trace] (6) at (-2.5, -0.75) {};
		\node [style=labelnode, inner sep=0.1 mm] (7) at (2.5, 0) {$R^\singletonSet$};
		\node [style=trace] (8) at (5.5, 0) {};
		\node [style=labelnode] (9) at (7.5, 0) {$=$};
		\node [style=empty diagram] (10) at (10.5,0) {};
	\end{pgfonlayer}
	\begin{pgfonlayer}{edgelayer}
		\draw [style=-] (1) to (2);
		\draw [style=-] (3) to (4);
		\draw [style=-] (5) to (6);
		\draw [style=-] (7) to (8);
	\end{pgfonlayer}
\end{tikzpicture}
		\end{equation}
	On the subcategory $\classicalSubcategory{\CategoryC}$, this environment structure must coincide with the canonical one of $\RMatCategory{R}$.
\end{enumerate}
We refer to $\classicalSubcategory{\CategoryC}$ as \textbf{classical theory}, and to its objects and morphisms as \textbf{classical systems} and \textbf{processes}. As diagrammatic convention, we use dashed wires for classical systems, and solid wires for generic ones.
\end{definition}

\newpage

\noindent $R$-probabilistic theories come with a number of native features that are commonplace in the modelling of quantum protocols: it is possible to exert classical control, to define tests with classical outcomes, to marginalise over classical outcomes, to work with preparations and measurements, and to apply any kind of classical pre- and post-processing. Amongst the many mixed quantum-classical processes, we can consider Bell-type measurement scenarios. An $N$-party \textbf{Bell-type measurement scenario} in an $R$-probabilistic theory is a process in the following form, where the processes $B_1,...,B_N$ and the state $\rho$ are all normalised (recall that a process $f: \SpaceH \rightarrow \SpaceK$ is said to be \textbf{normalised} if $\trace{\SpaceK} \circ f = \trace{\SpaceH}$):
\begin{equation}\label{BellTest}
\resizebox{!}{1.4cm}{\begin{tikzpicture}
	\begin{pgfonlayer}{nodelayer}
		\node [style=none] (0) at (-2, 1) {};
		\node [style=none] (1) at (-2, -1) {};
		\node [style=none, doubled] (2) at (0.5, 2.5) {};
		\node [style=none] (3) at (2.5, -2.5) {};
		\node [style=none, doubled] (4) at (0.5, -2.5) {};
		\node [style=none] (5) at (2.5, 2.5) {};
		\node [style=box] (6) at (0, 2.5) {$B_1$};
		\node [style=box] (7) at (0, -2.5) {$B_{N}$};
		\node [style=tightlabelnode] (8) at (0, 0) {$\goodvdots$};
		\node [style=tightlabelnode] (9) at (3.25, 2.5) {\small{$O_1$}};
		\node [style=tightlabelnode] (10) at (3.25, -2.5) {\small{$O_N$}};
		\node [style=tightlabelnode] (11) at (3.25, 0) {$\goodvdots$};
		\node [style=none] (12) at (-4.25, 2.75) {};
		\node [style=none] (13) at (-4.25, -2.75) {};
		\node [style=tightlabelnode] (14) at (-5, -2.75) {\small{$M_N$}};
		\node [style=tightlabelnode] (15) at (-5, 2.75) {\small{$M_1$}};
		\node [style=stateLarge] (16) at (-2.75, 0) {$\rho$};
		\node [style=tightlabelnode] (17) at (-5, 0) {$\goodvdots$};
		\node [style=none, doubled] (18) at (-0.5, -2.75) {};
		\node [style=none, doubled] (19) at (-0.5, 2.75) {};
	\end{pgfonlayer}
	\begin{pgfonlayer}{edgelayer}
		\draw [style=dashed] (2.center) to (5.center);
		\draw [style=dashed] (4.center) to (3.center);
		\draw [style=dashed, line width=5 pt, draw=white, in=180, out=0] (1.center) to (7);
		\draw [style=dashed, line width=5 pt, draw=white, in=180, out=0] (0.center) to (6);
		\draw [style=-, in=180, out=0] (1.center) to (7);
		\draw [style=-, in=180, out=0] (0.center) to (6);
		\draw [style=dashed] (12.center) to (19.center);
		\draw [style=dashed] (13.center) to (18.center);
	\end{pgfonlayer}
\end{tikzpicture}}
\end{equation} 
In the context of a Bell-type measurement scenario, the processes $B_1,...,B_N$ are often referred to as \textbf{measurements}, their inputs as \textbf{measurement choices} and their outputs as \textbf{(measurement) outcomes}. The following result from Ref. \cite{Gogioso2017a} ensures that non-locality in $R$-probabilistic theories can always be studied using the well-established sheaf-theoretic framework for non-locality and contextuality \cite{Abramsky2011}.
\begin{theorem}[\textbf{Bell-type measurement scenarios \cite{Gogioso2017a}}]\hfill\\
\label{thm_BellTestEM}
A Bell-type measurement scenario in an $R$-probabilistic theory always corresponds to a no-signalling empirical model in the sheaf-theoretic framework for non-locality and contextuality \cite{Abramsky2011}.
\end{theorem}
\noindent An immediate consequence of the connection with the sheaf-theoretic framework is that we can straightforwardly adapt a proof of Ref. \cite{Abramsky2011} to rule out non-locality in a large class of toy models.  
\begin{theorem}[\textbf{Locality of $R$-probabilistic theories over fields}] \hfill\\
\label{thm_localityFields}
If $R$ is a field, then all $R$-probabilistic theories are local. 
\end{theorem}\vspace{-0.3cm}

\section{Theories of wavefunctions valued in commutative involutive semirings}
\label{section_quantumLikeTheories}
\vspace{-0.2cm}

Note that two different linear structures intervene in the definition of quantum theory: the $\complexs$-linear structure of wavefunctions, modelling superposition, interference and phases, and the $\reals^+$-linear structure of probability distributions over classical systems. We have already seen that the framework of $R$-probabilistic theories replaces the probability semiring $\reals^+$ with a more general commutative semiring $R$ as a model of classical non-determinism. In this Section, we construct a large class of toy models of quantum theory by considering theories of wavefunctions with amplitudes valued in some commutative semiring $S$ with involution, generalising the field with involution $\complexs$ traditionally used in quantum mechanics. To do so, we consider the dagger compact category $\RMatCategory{S}$ (dagger and compact closed structure will be defined using the involution), and we require classical non-determinism to arise via the Born rule, as embodied by the CP* construction. The corresponding quantum-classical theory will therefore be modelled by $\CPStarCategory{\RMatCategory{S}}$, and the main result of this Section (Theorem \ref{thm_RprobabilisticCPStarCategories}) will show that this is an $R$-probabilistic theory (where $R$ is the sub-semiring of positive elements of $S$; see Definition \ref{def_semiringPositiveElements} below).

The category $\RMatCategory{S}$ for a commutative semiring $S$ with involution is defined as in the previous Section, but it comes with additional structure. Indeed, we can defined dagger and compact closed structures on $\RMatCategory{S}$ exactly as done in $\fdHilbCategory$ (which is $\RMatCategory{\complexs}$), with conjugation taken using the involution $^\ast$ of $S$ in place of complex conjugation. Each object $S^X$ in $\RMatCategory{S}$ comes with at least one orthonormal basis $\ket{x}_{x \in X}$, as well as an associated special commutative $\dagger$-Frobenius algebra $\hbox{\begin{tikzpicture} [scale=1.2,transform shape] %% DO NOT CHANGE

\def\deltax{0.3} %% CAN BE CHANGED
\def\deltay{0.5} %% DO NOT CHANGE

%\path[use as bounding box] (-\deltax,-\deltay) rectangle (\deltax,\deltay);

\node [dot, fill=\Zbwcolour] (mult) at (0,0) {};
%\node (mult_label_out) at (0,+\deltay) {};
%\draw[-] (mult) to (mult_label_out);

%\draw (current bounding box.south west) rectangle (current bounding box.north east);
\end{tikzpicture}}\!\!_X$:
\begin{equation}
\begin{tikzpicture}[scale=0.8]
	\begin{pgfonlayer}{nodelayer}
		\node [style=Zbwdot] (0) at (-14.5, 0) {};
		\node [style=Zbwdot] (1) at (7.25, 0) {};
		\node [style=tightlabelnode] (2) at (-16.5, 0) {};
		\node [style=tightlabelnode] (3) at (-12.5, -0.75) {};
		\node [style=tightlabelnode] (4) at (-12.5, 0.75) {};
		\node [style=tightlabelnode] (5) at (5.25, 0) {};
		\node [style=tightlabelnode] (6) at (-11, 0) {$=$};
		\node [style=none] (7) at (-15.25, -0.5) {\tiny$X$};
		\node [style=none] (8) at (8, -0.5) {\tiny$X$};
		\node [style=tightlabelnode] (9) at (9.5, 0) {$=$};
		\node [style=tightlabelnode] (10) at (-6, -0.25) {$\sum\limits_{x \in X} \ket{x} \otimes \ket{x} \otimes \bra{x}$};
		\node [style=tightlabelnode] (11) at (12, -0.25) {$\sum\limits_{x \in X} \bra{x}$};
	\end{pgfonlayer}
	\begin{pgfonlayer}{edgelayer}
		\draw [style=-] (2) to (0);
		\draw [style=-, in=-45, out=180] (3) to (0);
		\draw [style=-, in=180, out=45] (0) to (4);
		\draw [style=-] (5) to (1);
	\end{pgfonlayer}
\end{tikzpicture}
\end{equation}
For any group structure $G = (X,\cdot,1)$ on any finite set $X$, one also obtains an associated $\dagger$-Frobenius algebra $\hbox{\begin{tikzpicture} [scale=1.2,transform shape] %% DO NOT CHANGE

\def\deltax{0.3} %% CAN BE CHANGED
\def\deltay{0.5} %% DO NOT CHANGE

%\path[use as bounding box] (-\deltax,-\deltay) rectangle (\deltax,\deltay);

\node [dot, fill=\Dcolour] (mult) at (0,0) {};
%\node (mult_label_out) at (0,+\deltay) {};
%\draw[-] (mult) to (mult_label_out);

%\draw (current bounding box.south west) rectangle (current bounding box.north east);
\end{tikzpicture}}\!\!_G$ on $S^X$ by linearly extending the group multiplication and unit:
\begin{equation}
\begin{tikzpicture}[scale=0.8]
	\begin{pgfonlayer}{nodelayer}
		\node [style=Xbwdot] (0) at (-15, 0) {};
		\node [style=Xbwdot] (1) at (6, 0) {};
		\node [style=tightlabelnode] (2) at (-13, 0) {};
		\node [style=tightlabelnode] (3) at (-17, -0.75) {};
		\node [style=tightlabelnode] (4) at (-17, 0.75) {};
		\node [style=tightlabelnode] (5) at (8, 0) {};
		\node [style=tightlabelnode] (6) at (-11.5, 0) {$=$};
		\node [style=none] (7) at (-14.25, -0.5) {\tiny$G$};
		\node [style=none] (8) at (5.25, -0.5) {\tiny$G$};
		\node [style=tightlabelnode] (9) at (9.5, 0) {$=$};
		\node [style=tightlabelnode] (10) at (-6, -0.25) {$\sum\limits_{x,y \in X} \ket{x\cdot y} \otimes \bra{x} \otimes \bra{y}$};
		\node [style=tightlabelnode] (11) at (11.5, 0) {$\ket{1}$};
	\end{pgfonlayer}
	\begin{pgfonlayer}{edgelayer}
		\draw [style=-] (2) to (0);
		\draw [style=-, in=-135, out=0] (3) to (0);
		\draw [style=-, in=0, out=135] (0) to (4);
		\draw [style=-] (5) to (1);
	\end{pgfonlayer}
\end{tikzpicture}
\end{equation}
The $\dagger$-Frobenius algebra is commutative if and only if the group is, and it always satisfies the following:
\begin{equation}
\begin{tikzpicture}[scale=0.8]
	\begin{pgfonlayer}{nodelayer}
		\node [style=Xbwdot] (0) at (-5.5, 0) {};
		\node [style=tightlabelnode] (1) at (-7.5, 0) {};
		\node [style=Xbwdot] (2) at (-3.5, 0) {};
		\node [style=Xbwdot] (3) at (-3.5, 0) {};
		\node [style=tightlabelnode] (4) at (0, 0) {$=$};
		\node [style=none] (5) at (-6.25, -0.5) {\tiny$G$};
		\node [style=none] (6) at (-2.75, -0.5) {\tiny$G$};
		\node [style=tightlabelnode] (7) at (-1.5, 0) {};
		\node [style=Xbwdot] (8) at (-3.5, 0) {};
		\node [style=tightlabelnode] (9) at (6.5, 0) {};
		\node [style=tightlabelnode] (10) at (1.5, 0) {};
		\node [style=tightlabelnode] (11) at (4, 0.75) {$|G|$};
	\end{pgfonlayer}
	\begin{pgfonlayer}{edgelayer}
		\draw [style=-] (1) to (0);
		\draw [style=-, in=-45, out=-135] (2) to (0);
		\draw [style=-, in=135, out=45] (0) to (3);
		\draw [style=-] (7) to (8);
		\draw [style=-] (10) to (9);
	\end{pgfonlayer}
\end{tikzpicture}
\end{equation}
Unfortunately, $\hbox{\begin{tikzpicture} [scale=1.2,transform shape] %% DO NOT CHANGE

\def\deltax{0.3} %% CAN BE CHANGED
\def\deltay{0.5} %% DO NOT CHANGE

%\path[use as bounding box] (-\deltax,-\deltay) rectangle (\deltax,\deltay);

\node [dot, fill=\Dcolour] (mult) at (0,0) {};
%\node (mult_label_out) at (0,+\deltay) {};
%\draw[-] (mult) to (mult_label_out);

%\draw (current bounding box.south west) rectangle (current bounding box.north east);
\end{tikzpicture}}\!\!_G$ is not quasi-special (a.k.a. normalisable) unless the scalar $|G|$ takes the form $z_G^\ast z_G$ for some $z_G \in S$ which is multiplicatively invertible: when this is the case, however, we have a legitimate strongly complementary pair $(\hbox{\begin{tikzpicture} [scale=1.2,transform shape] %% DO NOT CHANGE

\def\deltax{0.3} %% CAN BE CHANGED
\def\deltay{0.5} %% DO NOT CHANGE

%\path[use as bounding box] (-\deltax,-\deltay) rectangle (\deltax,\deltay);

\node [dot, fill=\Zbwcolour] (mult) at (0,0) {};
%\node (mult_label_out) at (0,+\deltay) {};
%\draw[-] (mult) to (mult_label_out);

%\draw (current bounding box.south west) rectangle (current bounding box.north east);
\end{tikzpicture}}\!\!_X,\hbox{\begin{tikzpicture} [scale=1.2,transform shape] %% DO NOT CHANGE

\def\deltax{0.3} %% CAN BE CHANGED
\def\deltay{0.5} %% DO NOT CHANGE

%\path[use as bounding box] (-\deltax,-\deltay) rectangle (\deltax,\deltay);

\node [dot, fill=\Dcolour] (mult) at (0,0) {};
%\node (mult_label_out) at (0,+\deltay) {};
%\draw[-] (mult) to (mult_label_out);

%\draw (current bounding box.south west) rectangle (current bounding box.north east);
\end{tikzpicture}}\!\!_G)$ in $\RMatCategory{S}$ corresponding to the finite group $G$. When $G$ is abelian these strongly complementary pairs can be used (under additional constraints) to implement quantum protocols such as the algorithm to solve the abelian Hidden Subgroup Problem \cite{Vicary2012a,Gogioso2017b} or generalised Mermin-type arguments \cite{Gogioso2015,Gogioso2017c}. This also means that certain objects in $\RMatCategory{S}$ support fragments of the ZX calculus\footnote{To be precise, they always support the spider rules (but cups/caps for the two algebras are generally distinct), the bialgebra rules, the Hopf laws (with non-trivial antipode), the copy rules, and a generalised version of the $\pi$-copy rules (see Ref. \cite{Backens2014}). A Hadamard unitary can be defined if and only if the $S$-valued unitary multiplicative characters for $G$ form a basis for $S^X$, and in this case the colour-change rules are also supported (taking care to consider Hadamard adjoints where relevant).} \cite{Coecke2011,Backens2014}, opening the way to the application of well-established diagrammatic techniques.

In quantum theory, the probabilistic semiring $\reals^+$ arises as a sub-semiring of $\complexs$ fixed by complex conjugation, namely the sub-semiring of those elements $z \in \complexs$ taking the form $z = x^\ast x$: this is, essentially, a hallmark of the Born rule. In general commutative semirings with involution, elements in the form $x^\ast x$ need not be closed under addition, but it is true their closure under addition always form a semiring.
\begin{definition}\label{def_semiringPositiveElements}
Let $S$ be a commutative semiring with involution. Then we define its \textbf{sub-semiring of positive elements} $R$ to be the closure under addition in $S$ of the set $\suchthat{x^\ast x}{ x \in S}$.
\end{definition}

\noindent When classical non-determinism is introduced via the Born rule, quantum theory naturally gives rise to a probabilistic theory. Similarly, it is possible to prove that any theory of wavefunctions valued in a commutative semiring $S$ with involution gives rise to an $R$-probabilistic theory, where $R$ is the corresponding sub-semiring of positive elements. 
\begin{theorem}\label{thm_RprobabilisticCPStarCategories}
Let $S$ be a commutative semiring with involution, and let $R$ be its sub-semiring of positive elements. Then $\CPStarCategory{\RMatCategory{S}}$ is $R$-probabilistic under the $\CMonCategory$-enrichment inherited from $\RMatCategory{S}$.
\end{theorem}
Note that the scalars of $\CPStarCategory{\RMatCategory{S}}$ are the elements of $R$, and that the pure scalars are those in the form $\xi^\ast \xi$ for some $\xi \in S$: as a consequence, not all scalars of $\CPStarCategory{\RMatCategory{S}}$ need be pure (in contrast to what happens with ordinary quantum theory). In what follows, we will try as much as possible to construct theories where all scalars are pure, but there are examples (such as the case of $p$-adic quantum theory) where this cannot be achieved. When all scalars are pure, the requirement that $|G| = z_G^\ast z_G$ is always automatically satisfied for all finite groups $G$, and we only need to care about $|G|$ being invertible as a scalar in $S$ (a fact which always holds true whenever $S$ is a semi-field/field and $|G|$ is non-zero in $S$).

We will now proceed to construct a number of toy models within this framework. For each toy model we will try to: (i) study the phase group; (ii) discuss which Hidden Subgroup Problems can be efficiently solved, and which generalised Mermin-type arguments can be implemented, using the phase group and the strongly complementary pairs; (iii) assess the overall locality of the theory, via the connection with the sheaf-theoretic framework for non-locality and contextuality.

\newpage
\section{Real quantum theory}
\label{section_realQT}

The simplest non-conventional example is given by the ring $\reals$ of signed reals (with the trivial involution), which yields the \textbf{probability semiring} $\reals^+$ as its sub-semiring of positive elements; in particular, all positive elements are pure scalars. The corresponding probabilistic theory $\CPStarCategory{\RMatCategory{\reals}}$ is known as \textbf{real quantum theory} \cite{Jordan1934,Baez2012,Belenchia2012,Wilce2016}: it is arguably the most well-studied of the quantum-like theories, and the closest to ordinary quantum theory. Thus said, real quantum theory can be distinguished from ordinary quantum theory because it fails to be \textit{locally tomographic} \cite{Araki1980,Wootters1990,Chiribella2010}, i.e. bipartite (mixed) states in real quantum theory cannot in general be distinguished by product measurements alone. Equivalently, one can check that the CP maps $\CPMdoubled{\sigma_x} + \CPMdoubled{\sigma_z} - \CPMdoubled{\id{\reals^2}}$ and $\CPMdoubled{\sigma_y}$ on $\reals^2$ in $\CPMCategory{\RMatCategory{\reals}}$ cannot be distinguished by applications to mixed states of $\reals^2$ alone, because the latter are described by density matrices which are always real symmetric\footnote{By $\sigma_x$, $\sigma_y$ and $\sigma_z$ we have denoted the complex qubit Pauli matrices, which give rise to real CP maps on $\reals^2$ when doubled.}.

The group of phases in $\reals$ is $\{\pm 1\} \isom \integersMod{2}$, and non-trivial interference is possible in real quantum theory. For example, each of the Pauli $X$ eigenstates $\ket{\pm} := \frac{1}{\sqrt{2}}(\ket{0}\pm\ket{1})$ of the qubit $\reals[\integersMod{2}]$ in real quantum theory yields the uniform distribution when measured in the Pauli $Z$ basis $\ket{0},\ket{1}$, but their superposition $\frac{1}{\sqrt{2}}(\ket{+}+\ket{-})$ yields the outcome corresponding to $\ket{0}$ with certainty. Simon's problem, and other Hidden Subgroup Problems on $\integersMod{2}^N$, can be solved efficiently in real quantum theory.
More generally, consider a classical structure $\hbox{\begin{tikzpicture} [scale=1.2,transform shape] %% DO NOT CHANGE

\def\deltax{0.3} %% CAN BE CHANGED
\def\deltay{0.5} %% DO NOT CHANGE

%\path[use as bounding box] (-\deltax,-\deltay) rectangle (\deltax,\deltay);

\node [dot, fill=\Zbwcolour] (mult) at (0,0) {};
%\node (mult_label_out) at (0,+\deltay) {};
%\draw[-] (mult) to (mult_label_out);

%\draw (current bounding box.south west) rectangle (current bounding box.north east);
\end{tikzpicture}}\!\!$ on $\reals^d$ with enough classical states, which corresponds to an orthonormal basis of $\reals^d$ (because $\reals$ is multiplicatively cancellative \cite{Coecke2013b}). Then the group of $\hbox{\begin{tikzpicture} [scale=1.2,transform shape] %% DO NOT CHANGE

\def\deltax{0.3} %% CAN BE CHANGED
\def\deltay{0.5} %% DO NOT CHANGE

%\path[use as bounding box] (-\deltax,-\deltay) rectangle (\deltax,\deltay);

\node [dot, fill=\Zbwcolour] (mult) at (0,0) {};
%\node (mult_label_out) at (0,+\deltay) {};
%\draw[-] (mult) to (mult_label_out);

%\draw (current bounding box.south west) rectangle (current bounding box.north east);
\end{tikzpicture}}\!\!$-phase gates is isomorphic to the group $\integersMod{2}^{d-1}$ of $(d-1)$-bit strings under bitwise xor. Because of the structure of phase groups, generalised Mermin-type arguments only yield local empirical models \cite{Gogioso2015}. Nevertheless, Bell's theorem goes through in real quantum theory (as it only involves states and measurements on the $ZX$ great circle of the Bloch sphere), which is therefore a non-local probabilistic theory.

\section{Relational quantum theory}
\label{section_relationalQT}

Examples of an entirely different nature are given by considering distributive lattices $\Omega$ (with the trivial involution), which yield themselves back as their sub-semirings of positive elements (because of multiplicative idempotence); in particular, all positive elements are pure scalars. Distributive lattices seem to be almost as far as one can get from the probabilistic semiring $\reals^+$, but the category $\RMatCategory{\Omega}$ has been studied extensively as a toy model for quantum theory (especially in the boolean case $\Omega = \mathbb{B}$) \cite{Pavlovic2009,Abramsky2012,Evans2009,Zeng2015,Coecke2012a}, and the corresponding CPM category has also received some attention on its own \cite{Marsden,Gogioso2015g}. We refer to the corresponding $\Omega$-probabilistic (or \textbf{possibilistic}) theory as \textbf{relational quantum theory}.

The group of phases in $\Omega$ is the singleton $\{1\}$, and no interference is possible in relational quantum theory. Relational quantum theories also feature very few quantum-to-classical transitions: there is a unique basis on each system, namely the one given by the elements of the underlying set. They are local tomographic on pure states, but they fail to be tomographic altogether on mixed states: for example, the pure state $\ket{\psi}\bra{\psi}$ for $\ket{\psi} := \ket{0}+\ket{1}$ and the mixed state $\ket{0}\bra{0} + \ket{1}\bra{1}$ are distinct, but cannot be distinguished by measurement. In fact, a characteristic trait of relational quantum theories is exactly that superposition and mixing are essentially indistinguishable (because of idempotence) \cite{Abramsky2012,Marsden,Gogioso2015g}. Classical structures $\hbox{\begin{tikzpicture} [scale=1.2,transform shape] %% DO NOT CHANGE

\def\deltax{0.3} %% CAN BE CHANGED
\def\deltay{0.5} %% DO NOT CHANGE

%\path[use as bounding box] (-\deltax,-\deltay) rectangle (\deltax,\deltay);

\node [dot, fill=\Zbwcolour] (mult) at (0,0) {};
%\node (mult_label_out) at (0,+\deltay) {};
%\draw[-] (mult) to (mult_label_out);

%\draw (current bounding box.south west) rectangle (current bounding box.north east);
\end{tikzpicture}}\!\!$ in relational quantum theory over the booleans are known to correspond to abelian groupoids $\oplus_{i\in I} G_i$ \cite{Pavlovic2009}, and the corresponding group of $\hbox{\begin{tikzpicture} [scale=1.2,transform shape] %% DO NOT CHANGE

\def\deltax{0.3} %% CAN BE CHANGED
\def\deltay{0.5} %% DO NOT CHANGE

%\path[use as bounding box] (-\deltax,-\deltay) rectangle (\deltax,\deltay);

\node [dot, fill=\Zbwcolour] (mult) at (0,0) {};
%\node (mult_label_out) at (0,+\deltay) {};
%\draw[-] (mult) to (mult_label_out);

%\draw (current bounding box.south west) rectangle (current bounding box.north east);
\end{tikzpicture}}\!\!$-phase gates is isomorphic to $\prod_{i \in I} G_i$. It can be shown that generalised Mermin-type arguments only yield local empirical models \cite{Gogioso2015}. In fact, it can be shown that that relational quantum theories are entirely local \cite{Abramsky2012,Gogioso2015g}.

\section{Hyperbolic quantum theory}
\label{section_hyperbolicQT}

Turning our attention back to real algebras, we can consider the commutative ring of \textbf{split complex numbers} $\splitComplexs := \reals[X]/(X^2-1)$, a two-dimensional real algebra. Split complex numbers take the form $(x+j y)$, where $x,y \in \reals$ and $j^2 = 1$; in particular, they have non-trivial zero-divisors in the form $a(1\pm j)$, because $(1+j)(1-j)=1-j^2 = 0$. They come with the involution $(x + j y)^\ast := x - jy$, which yields the \textbf{signed-probability ring} $\reals$ as sub-semiring of positive elements; in particular, all positive elements are pure scalars. We refer to the corresponding $\reals$-probabilistic theory $\CPStarCategory{\RMatCategory{\splitComplexs}}$ as \textbf{hyperbolic quantum theory}\footnote{Clifford referred to functions of split complex numbers as \inlineQuote{functions of a motor variable} \cite{Clifford1871}, so we could say that hyperbolic quantum theory is the theory of \textbf{wavefunctions of a motor variable} (how does \textbf{motor quantum theory} sound?).} \cite{Khrennikov2003,Khrennikov2010,Nyman2011}. 

Hyperbolic quantum theory is an extremely interesting theory. On the one hand, it contains real quantum theory as a sub-theory\footnote{By which we mean that $\splitComplexs$ contain $\reals$ as a sub-ring fixed by the involution.}, and as a consequence every scenario and protocol which can be implemented in real quantum theory (such as the algorithm to efficiently solve Simon's problem \cite{Gogioso2017b}) can also be implemented in hyperbolic quantum theory. On the other hand, hyperbolic quantum theory is a local theory, in the sense that every empirical model arising in hyperbolic quantum theory admits a local hidden variable model in terms of signed probabilities (the notion of classical non-determinism for hyperbolic quantum theory) \cite{Abramsky2011}. While signed probabilities might at first sound unphysical, an operational interpretation exists in terms of unsigned probabilities on signed events \cite{Abramsky2014}\footnote{Where the sign of the events themselves cannot be observed, yielding an epistemic restriction which could be seen as not-too-far-removed from the one which originally motivated Spekkens's toy model \cite{Spekkens2007,Catani2017}}. 

The group of phases in $\splitComplexs$ consists of the elements with square norm $1$, i.e. the elements in the form $x+jy$ which lie on the following unit hyperbola of the real plane:
\begin{equation}
1=(x+jy)^\ast(x+jy)=x^2-y^2
\end{equation} 
In fact, the natural geometry for the split complex numbers is that of the real plane endowed with the Lorentzian metric $-dy^2+dx^2$, i.e. that of the Minkowski plane. Just like multiplication by phases in $\complexs$ forms the circle group $U(1)$ of rotations around the origin for the Euclidean plane, multiplication by phases in $\splitComplexs$ forms the group $SO(1,1)$ of orthochronous homogeneous Lorentz transformations for the Minkowski plane. We have the isomorphism of Lie groups $\integersMod{2} \times \reals \isom SO(1,1)$ given by $(s,\theta) \mapsto s (\cosh(\theta)+j\sinh(\theta))$: as a consequence, the $\splitComplexs$-valued multiplicative characters for finite groups are exactly the same as the $\reals$-valued multiplicative characters, and the only finite groups with enough multiplicative characters to form a Fourier basis are the ones in the form $\integersMod{2}^N$; Simon's problem, and other Hidden Subgroup Problems for $\integersMod{2}^N$, can be efficiently solved in hyperbolic quantum theory, despite the latter being local. Things are different for infinite groups such as $\integers$, which have enough $\splitComplexs$-valued multiplicative characters but not enough $\reals$-valued multiplicative characters.

Now consider a classical structure $\hbox{\begin{tikzpicture} [scale=1.2,transform shape] %% DO NOT CHANGE

\def\deltax{0.3} %% CAN BE CHANGED
\def\deltay{0.5} %% DO NOT CHANGE

%\path[use as bounding box] (-\deltax,-\deltay) rectangle (\deltax,\deltay);

\node [dot, fill=\Zbwcolour] (mult) at (0,0) {};
%\node (mult_label_out) at (0,+\deltay) {};
%\draw[-] (mult) to (mult_label_out);

%\draw (current bounding box.south west) rectangle (current bounding box.north east);
\end{tikzpicture}}\!\!$ corresponding to an orthonormal basis of $\splitComplexs^d$ (we have to ask explicitly for orthogonality, because the result of \cite{Coecke2013b} does not apply to hyperbolic quantum theory: $\splitComplexs$ has non-trivial zero-divisors, and hence it is not multiplicatively cancellative). The group of $\hbox{\begin{tikzpicture} [scale=1.2,transform shape] %% DO NOT CHANGE

\def\deltax{0.3} %% CAN BE CHANGED
\def\deltay{0.5} %% DO NOT CHANGE

%\path[use as bounding box] (-\deltax,-\deltay) rectangle (\deltax,\deltay);

\node [dot, fill=\Zbwcolour] (mult) at (0,0) {};
%\node (mult_label_out) at (0,+\deltay) {};
%\draw[-] (mult) to (mult_label_out);

%\draw (current bounding box.south west) rectangle (current bounding box.north east);
\end{tikzpicture}}\!\!$-phase gates is isomorphic to $(\integersMod{2}\times \reals)^{d-1}$, and has $\integersMod{2}^{d-1}$ as a maximal finite subgroup: as a consequence, generalised Mermin-type arguments (which involve finite groups) only yield local empirical models, just as in real quantum theory. However, future extensions of Mermin-type arguments to infinite groups might yield different results: this is because subgroups like $(\{0\}\times\integers) \trianglelefteq (\integersMod{2} \times \reals)$ would become available, and there are equations (such as $2\theta=1$) which have no solutions in the subgroup $\{0\} \times \integers$ but have solutions (e.g. $\theta = (0,\frac{1}{2})$ and  $\theta = (1,\frac{1}{2})$) in the larger group $\integersMod{2} \times \reals$.

\section{Parity quantum theory}
\label{section_parityQT}

A simple variation on relational quantum theory (over the booleans) is given by using symmetric difference of sets, instead of union, as the superposition operation. This leads us to consider the finite field with two elements $\integersMod{2}:=(\{0,1\},+,0,\times,1)$, with trivial involution, in place of the booleans $\mathbb{B}:=(\{0,1\},\vee,0,\times,1)$, also with trivial involution. The multiplication is the same, but addition is now non-idempotent, and superposition is no longer the same as mixing. The \textbf{parity semiring} $\integersMod{2}$ yields itself back as its sub-semiring of positive elements (in particular, all positive elements are pure scalars), and we refer to the corresponding $\integersMod{2}$-probabilistic theory $\CPStarCategory{\RMatCategory{\integersMod{2}}}$ as \textbf{parity quantum theory}. 

\begin{remark}
Parity quantum theory as defined here (the same as in Ref. \cite{DeBeaudrap2014}) pretty much coincides with the $\integersMod{2}$ case of modal quantum theory \cite{Schumacher2012,Schumacher2016}, but it should be noted that the philosophical interpretation of $\integersMod{2}$-valued probabilities is significantly different. In modal quantum theory, the interest is in generating possibilistic tables by using finite fields, subsequently interpreting all zero values as the boolean $0$ and all non-zero values as the boolean $1$. In parity quantum theory, the non-determinism itself is interpreted to be natively $\integersMod{2}$-valued, and no attempt is made to translate the resulting empirical models into possibilistic ones. Indeed, such an interpretation would not be natural within our semiring-oriented framework, as no semiring homomorphism can exists from any finite field into the booleans. 
\end{remark}

The group of phases in $\integersMod{2}$ is the singleton $\{1\}$, but interference is still possible in parity quantum theory: this somewhat counter-intuitive situation is made possible by the fact that $1$ is its own additive inverse in $\integersMod{2}$, so that triviality of the group of phases is slightly deceptive. Indeed, consider the four two-qubit states below, which form an orthonormal basis for $\integersMod{2}^2$:
\begin{align}
\ket{\psi_{012}} := \ket{00}+\ket{01}+\ket{10} \hspace{2cm} \ket{\psi_{123}} := \ket{01}+\ket{10}+\ket{11} \nonumber \\
\ket{\psi_{230}} := \ket{10}+\ket{11}+\ket{00} \hspace{2cm} \ket{\psi_{301}} := \ket{11}+\ket{00}+\ket{01}
\end{align}
For example, we have $\ket{10} = \ket{\psi_{012}}+\ket{\psi_{123}}+\ket{\psi_{230}}$. When measured in the computational basis $\ket{00}, \ket{01}, \ket{10}, \ket{11}$, the normalised states $\ket{01}$, $\ket{10}$ and $\ket{\psi_{012}}$ all have non-zero $\integersMod{2}$-probability of yielding an outcome in the set $\{01,10\}$, but their superposition $\ket{01}+\ket{10}+\ket{\psi_{012}} = \ket{00}$ (also a normalised state) has zero $\integersMod{2}$-probability of yielding an outcome in that set. 

Because the group of phases is trivial, so are all the groups of phase gates, as well as all the $\integersMod{2}$-valued multiplicative characters of all groups; as a consequence, parity quantum theory admits no non-trivial generalised Mermin-type arguments, and no implementation of the algorithm to solve the HSP. Furthermore, Theorem \ref{thm_localityFields} shows that parity quantum theory is local, because $\integersMod{2}$ is a field.

$R$-probabilistic theories can be similarly constructed for modal quantum theory  over any other finite field $\finiteField{p^n}$ \cite{Schumacher2012,Schumacher2016}, by taking $S := \finiteField{p^n}$ with the trivial involution. However, these theories have a lot of non-pure scalars---namely the $(p^n-1)/2$ non-squares in $\finiteField{p^n}$---and their phases are close to trivial---namely they are $\{\pm 1\}$ if $p>2$ and $\{1\}$ if $p=2$. Instead, we will consider a more sophisticated construction based on quadratic extensions of finite fields, which we will refer to as \inlineQuote{finite-field quantum theory}.\footnote{A related construction features in Ref. \cite{DeBeaudrap2014}, but from a computational complexity angle rather than a physical theory one.} %Contrary to modal quantum theory, finite-field quantum theory comes with a large group of phases, allowing non-trivial implementations of quantum protocols such as the algorithm for the abelian HSP and Mermin-type non-locality arguments; at the same time, it is an entirely local theory---a consequence of the finite-field interpretation of its classical non-determinism, contrasting with the possibilistic interpretation given in modal quantum theory. As such, it will provide a new interesting family of local toy models (alternative to Spekkens's toy model \cite{Spekkens2007,Catani2017,Disilvestro2016}) in which to implement non-trivial quantum protocols within the framework of Categorical Quantum Mechanics.     
Finite-field quantum theory is a local theory (by Theorem \ref{thm_localityFields}), in which it is nonetheless possible to formulate non-trivial quantum algorithms, as well as non-trivial Mermin-type \inlineQuote{non-locality} arguments. This is in stark contrast with more traditional toy models such as Spekkens's toy model \cite{Spekkens2007,Catani2017,Disilvestro2016} and relational quantum theory, in which the quantum Fourier transform cannot be performed for non-trivial groups \cite{Gogioso2015d} (precluding the implementation of algorithms based on it), and in which all Mermin-type arguments are necessarily trivial \cite{Coecke2010b,Coecke2012c,Gogioso2015}.

\section{Finite-field quantum theory}
\label{section_finitefieldQT}

Consider a finite field $\finiteField{p^n}$ (with $p$ odd), and let $\epsilon$ be a generator for the cyclic group $\finiteField{p^n}^\times$ of invertible (aka non-zero) elements in $\finiteField{p^n}$ (i.e. a primitive element for $\finiteField{p^n}$). We consider the ring $\finiteField{p^n}[\sqrt{\epsilon}] := \finiteField{p^n}[X^2-\epsilon]$, equipped with the involution $(x+y\sqrt{\epsilon})^\ast := (x-y\sqrt{\epsilon})$: because $\epsilon$ is a primitive element, $\finiteField{p^n}(\sqrt{\epsilon}) \isom \finiteField{p^{2n}}$ is a field. We are in fact working with the quadratic extension of fields $\finiteField{p^n}(\sqrt{\epsilon}) / \finiteField{p^n}$, equipped with the usual involution and (squared) norm from Galois theory:
\begin{equation}
\big|x+y\sqrt{\epsilon}\big|^2 = (x-y\sqrt{\epsilon})(x+y\sqrt{\epsilon}) = x^2 - \epsilon y^2
\end{equation}
The sub-field $\finiteField{p^n}$ (given by the elements in the form $x+0\sqrt{\epsilon}$) is the sub-semiring of positive elements (and we will shortly see that all positive elements are pure scalars). 

The phases in $\finiteField{p^{n}}(\sqrt{\epsilon})$ are the points $(x,y)$ of the $\finiteField{p^n}^2$ plane lying on the unit hyperbola $x^2 - \epsilon y^2 = 1$, which does not factor as a product of two lines because $\epsilon$ is a primitive element. The following iconic result of Galois theory, due to Hilbert, can be used to characterise them (see e.g. Ref. \cite{Hilbert1998} for a proof). 
\begin{theorem}[\textbf{Hilbert's Theorem 90}]\hfill\\
Let $L/K$ be a finite cyclic field extension, and let $\sigma: L \rightarrow L$ be a generator for its cyclic Galois group. Then the multiplicative group of elements $\xi \in L$ of unit relative norm $N_{L/K}(\xi)=1$ is isomorphic to the quotient group $L^\times / K^\times$.
\end{theorem}
\begin{corollary}\label{cor_phasesFFQT}
The phases in $\finiteField{p^{n}}(\sqrt{\epsilon})$ form the cyclic group $\finiteField{p^{2n}}^\times/\finiteField{p^n}^\times \isom \integersMod{p^n+1}$.
\end{corollary}
\noindent Another interesting consequence of Hilbert's Theorem 90 is the fact that the positive elements in finite-field quantum theory are all pure scalars.
\begin{lemma}\label{lem_purescalarsFFQT}
All scalars in $\CPStarCategory{\RMatCategory{\finiteField{p^{n}}(\sqrt{\epsilon})}}$ are pure.
\end{lemma}

We have seen that finite-field quantum theory comes with a non-trivial phase group, which in turn allows for non-trivial implementations of certain quantum protocols. We open with a result about the Quantum Fourier Transform, which combined with the main result of Ref. \cite{Gogioso2017b} implies that the Hidden Subgroup Problem can be solved efficiently in finite-field quantum theory for arbitrarily large families of finite abelian groups (as $p^n$ grows larger). 
\begin{lemma}\label{lem_HSPinFFQT}
Let $G$ be a finite abelian group. Then $G$ has enough $\finiteField{p^n}(\sqrt{\epsilon})$-valued unitary multiplicative characters if and only if $G \isom \prod_{k=1}^{K} \integersMod{p_k^{e_k}}$ with $p_k^{e_k} | p^n+1$ for all $k=1,...,K$. When this is the case, the Hidden Subgroup Problem for $G$ can be solved efficiently in finite-field quantum theory. 
\end{lemma}
Now consider a classical structure $\hbox{\begin{tikzpicture} [scale=1.2,transform shape] %% DO NOT CHANGE

\def\deltax{0.3} %% CAN BE CHANGED
\def\deltay{0.5} %% DO NOT CHANGE

%\path[use as bounding box] (-\deltax,-\deltay) rectangle (\deltax,\deltay);

\node [dot, fill=\Zbwcolour] (mult) at (0,0) {};
%\node (mult_label_out) at (0,+\deltay) {};
%\draw[-] (mult) to (mult_label_out);

%\draw (current bounding box.south west) rectangle (current bounding box.north east);
\end{tikzpicture}}\!\!$ with enough classical states on a $d$-dimensional quantum system in finite-field quantum theory, which corresponds to an orthonormal basis of the vector space $\big(\finiteField{p^n}(\sqrt{\epsilon})\big)^d$ (because $\finiteField{p^n}(\sqrt{\epsilon})$ is multiplicatively cancellative \cite{Coecke2013b}). Then the group of $\hbox{\begin{tikzpicture} [scale=1.2,transform shape] %% DO NOT CHANGE

\def\deltax{0.3} %% CAN BE CHANGED
\def\deltay{0.5} %% DO NOT CHANGE

%\path[use as bounding box] (-\deltax,-\deltay) rectangle (\deltax,\deltay);

\node [dot, fill=\Zbwcolour] (mult) at (0,0) {};
%\node (mult_label_out) at (0,+\deltay) {};
%\draw[-] (mult) to (mult_label_out);

%\draw (current bounding box.south west) rectangle (current bounding box.north east);
\end{tikzpicture}}\!\!$-phase gates in $\CPStarCategory{\RMatCategory{\finiteField{p^n}(\sqrt{\epsilon})}}$ is isomorphic to the group $\integersMod{p^n+1}^{d-1}$. %The following result then characterises which generalised Mermin-type arguments \cite{Gogioso2017c} can be implemented in finite-field quantum theory. 
\begin{lemma}\label{lem_merminFFQT}
It is possible to formulate non-trivial generalised Mermin-type arguments\footnote{By \textit{non-trivial} we mean arguments for systems of equation having no solutions in the subgroup of classical states.} in finite-field quantum theory if and only if $p^n+1$ is not a square-free natural number.
\end{lemma}

While finite-field quantum theory and parity quantum theory might not have as direct a physical interpretation as hyperbolic quantum theory and relational quantum theory, they come with the major advantage of having wavefunction valued over a field, so that objects are finite-dimensional vector spaces (equipped with a non-standard inner product, in the case of finite-field quantum theory). This opens the door for a systematic study of quantum systems in these theories using standard tools from finite geometry. Further investigation in this direction is left to future work.

\section{\texorpdfstring{$p$}{p}-adic quantum theory}
\label{section_pAdicQT}

We now look at the construction of \textbf{$p$-adic quantum mechanics} \cite{Vladimirov1989,Ruelle1989,Khrennikov1991,Khrennikov1993,Palmer2016,Palmer2016a}, where $R:=Q_p$ is the field of $p$-adic numbers, and $S$ is some quadratic extension. In this Section, we will use the notation $Q_p$ to denote the $p$-adic numbers, and $Z_p$ to denote the $p$-adic integers, to distinguish them from the finite field $\integersMod{p}$ of integers modulo $p$; note that this convention is different from the one used in many texts on $p$-adic arithmetic, where $\integersMod{p}$ is used for the $p$-adic integers (and $\rationals_p$ for the $p$-adic numbers). 

When $p > 2$, the $p$-adic numbers $x:= p^\ord{x}\sum_{i=0}^{+\infty} x_i p^i$ fall within four distinct quadratic classes---jointly labelled by the parity of the order $\ord{x} \in \integers$ and by the quadratic class of the first non-zero digit $x_0 \in \integersMod{p}^\times$---corresponding to three inequivalent quadratic extensions. This means that there is no way to obtain all positive elements as pure scalars by a single quadratic extension. This would seem to indicate that mixed states play a necessary role in the emergence of $p$-adic probabilities, which cannot all be obtained from pure states alone: the potential physical significance of this observation might become the topic of future work on $p$-adic quantum theory within CQM.

We consider the quadratic extension $S:=Q_p(\sqrt{\epsilon})$, where $p \geq 3$ and $\epsilon$ is a primitive element in the field $\integersMod{p}$ of integers modulo $p$, and we follow the presentation of Ref. \cite{Ruelle1989}. A generic element of $Q_p(\sqrt{\epsilon})$ takes the form $c + s \sqrt{\epsilon}$, for $c,s \in Q_p$, and its square norm is $|c+s\sqrt{\epsilon}|^2 = (c-s\sqrt{\epsilon})(c+s\sqrt{\epsilon}) = c^2 - \epsilon s^2$. Whether an element $x \in Q_p$ can be written in this form, i.e. whether is is a pure scalar in $\CPStarCategory{\RMatCategory{Q_p(\sqrt{\epsilon})}}$, is determined by the \textit{sign function} $\sgn{\epsilon}{x}$, which takes the value $+1$ if $x = c^2 - \epsilon s^2$ for some $c,s \in Q_p$, and the value $-1$ otherwise. An explicit form for the sign function (in the $p \neq 2$ case) is given by Equation (2.34) of Ref. \cite{Ruelle1989}, which specialised to our case ($\tau = \epsilon$ and $\ord{\tau} = 0$) reads $\sgn{\epsilon}{x} = (-1)^{\ord{x}}$. Hence the pure scalars in $\CPStarCategory{\RMatCategory{Q_p(\sqrt{\epsilon})}}$ are exactly the $p$-adic numbers $x$ with even order $\ord{x}$; closure of this set under addition yields $R := Q_p$ as sub-semiring (field, in fact) of positive elements in $S:= Q_p(\sqrt{\epsilon})$.

The phases in $p$-adic quantum theory are those $\xi := (c + s \sqrt{\epsilon}) \in Q_p(\sqrt{\epsilon})$ such that $\xi^\ast \xi = c^2 - \epsilon s^2 = 1$. In Ref. \cite{Ruelle1989} (Equation (4.35) of Section IV.C, and Equation (C12b) of Appendix C.3) it is shown that phases form a multiplicative group $C_\epsilon$ isomorphic to the additive group $\integersMod{p+1} \times p Z_p$---here $(\integersMod{p+1},+,0)$ are the integers modulo $p+1$, while $(p Z_p, +, 0)$ is the additive subgroup of $Z_p$ formed by those $p$-adic integers which are divisible by $p$. In particular, $C_\epsilon$ is an infinite group with the cardinality of the continuum, and each \inlineQuote{sheet} $pZ_p$ is a profinite\footnote{And hence both compact and totally disconnected.} torsion-free group, which is best understood by looking at the descending normal series $p Z_p \triangleright p^2 Z_p \triangleright ... \triangleright p^m Z_p \triangleright ...$ and considering the finite cyclic quotients  $p^n Z_p / p^m Z_p \isom \integersMod{p^{m-n}}$.

The scalar $|G|$ is always invertible, and it is in the form $|G| = z_G^\ast z_G$ if and only if the largest power of $p$ which divides $|G|$ is even. Furthermore, $G$ has enough $Q_p(\sqrt{\epsilon})$-valued multiplicative characters if and only if $G \isom \prod_{k=1}^{K} \integersMod{p_k^{e_k}}$ with $p_k^{e_k} | p+1$ for all $k=1,...,K$ (in the light of Hensel's Lemma, this parallelism between $p$-adic quantum theory and finite-field quantum theory on $\finiteField{p}$ should not come as a big surprise): finite abelian groups $G$ satisfying this condition admit efficient solutions for Hidden Subgroup Problems in $p$-adic quantum theory (because we necessarily have that $p$ cannot divide $|G|$). Similarly, it is possible to formulate non-trivial generalised Memrin-type arguments in $p$-adic quantum theory if and only if $p+1$ is not square-free. Thus said, $p$-adic quantum theory is a local theory by virtue of Theorem \ref{thm_localityFields}.

\begin{remark}
Similar considerations apply to the the construction of $p$-adic quantum theory for the other two quadratic extensions $Q_p(\sqrt{p})$ and $Q_p(\sqrt{p\epsilon})$ available in the case of $p \geq 3$ (although the cases $p=3$ and $p \geq 5$ have to be treated separately), as well as the seven quadratic extensions available in the case of $p=2$. The phase groups take a similar (but not identical) form to the one presented here, and the full details can be found in Ref. \cite{Ruelle1989} (Section IV.C and Appendices C.3, C.4).
\end{remark}

\section{Tropical quantum theory}
\label{section_tropicalQT}

Relational quantum theory involves semirings which are both additively and multiplicatively idempotent, parity quantum theory involves a semiring which is only multiplicatively idempotent, and ordinary quantum theory involves a semiring which is neither additively nor multiplicatively idempotent. We now give examples of theories with wavefunctions based in semirings which are additively idempotent but not multiplicatively idempotent, namely the tropical semirings \cite{Simon1988,Maslov1987,Simon1994,Pin1998,Speyer2009}. 
\begin{definition}
A \textbf{tropical semiring} is the commutative semiring $(M,\min,\infty,+,0)$ obtained from a totally ordered commutative monoid $(M,+,0,\leq)$ having an absorbing element $\infty$ which is larger than all elements in the monoid. In the tropical semiring, $\min$ is the addition, $\infty$ is the additive unit, $+$ is the multiplication and $0$ is the multiplicative unit. The nomenclature is extended to semirings isomorphic to the explicitly min-plus semirings used above (e.g. max-plus formulations, or the Viterbi semiring).
\end{definition}
Examples of tropical semirings appearing in the literature  include the tropical reals $(\reals\sqcup\{\infty\},\min,\infty,+,0)$, the tropical integers $(\integers\sqcup\{\infty\},\min,\infty,+,0)$, the tropical naturals $(\naturals\sqcup\{\infty\},\min,\infty,+,0)$, and the Viterbi semiring $([0,1],\max,0,\cdot,1)$ (which is a tropical semiring because it is isomorphic to the explicitly min-plus semiring $(\reals^+\sqcup\{\infty\},\min,\infty,+,0)$ via the semiring homomorphism $x \mapsto -\log x$). In fact, there is an easy characterisation of which commutative semirings arise as tropical semirings (the proof is omitted as it is a straightforward check).
\begin{lemma}
A commutative semiring $(S,+,0,\cdot,1)$ is a tropical semiring if and only if for all $a,b \in S$ we have $a = a+b$ or $b = a+b$ (in which case we can set $\min(a,b) = a+b$).
\end{lemma}
\noindent From now on, we will revert back to usual semiring notation, and we will rely on the result above to connect with the min-plus notation typical of tropical geometry \cite{Speyer2009}. We will, however, remember that tropical semirings come with a total order respected by the multiplication, and we will occasionally use $\min$, $\max$ and $\leq$ in addition to the addition/multiplication.

\begin{lemma}\label{lem_tropicalSemiringPositiveEls}
The only involution possible on a tropical semiring $(S,+,0,\cdot,1)$ is the trivial one, and the positive elements form the sub-semiring of squares $(\suchthat{x^2}{x\in S},+,0,\cdot,1)$.
\end{lemma}

\noindent If $S$ is a tropical semiring and $R:=(\suchthat{x^2}{x \in S},+,0,\cdot 1)$ is its sub-semiring of positive elements, we refer to the $R$-probabilistic theory $\CPStarCategory{\RMatCategory{S}}$ as \textbf{tropical quantum theory}.

Just as in the case of relational quantum theory, the group of phases in a tropical semiring $S$ is always trivial (because $x^2 = 1$ implies $x=1$ in any totally ordered monoid $(S,\cdot,1,\leq)$), and there is no interference. Similarly, there is a unique orthonormal basis on each system, the only unitaries/invertible maps are permutations, and superposition cannot be distinguished from mixing by measurements alone. Tropical quantum theory does not admit any implementation of the algorithm for the abelian Hidden Subgroup Problem, nor does it admit any generalised Mermin-type non-locality arguments.

The parallelisms with relational quantum theory become less surprising when one realises that tropical quantum theory is another generalisation of quantum theory over the booleans: the latter form a totally ordered distributive lattice, and hence are a particular case of tropical semiring. (Proof of the following result is omitted, as it is a straightforward check.)
\begin{lemma} 
Any totally ordered distributive lattice $(\Omega,\vee,\bot,\wedge,\top)$ is a tropical semiring $(\Omega,\wedge,\top,\vee,\bot)$; conversely, every tropical semiring $(S,+,0,\cdot,1)$ which has $1$ as least element and such that $x^2 = x$ for all $x \in S$ is a totally ordered distributive lattice $(S, \cdot,1,+,0)$.
\end{lemma}
\noindent In the light of the result above, we expect tropical quantum theory to be local, exactly like relational quantum theory, but further investigation of this question is left to future work.

\section{Conclusions and Future Work}
\label{section_conclusions}

In the first two Sections of this work, we have provided a general framework, based on enrichment of CP* categories, for the construction of toy models of quantum theory. Specifically, we have focussed our efforts on theories of wavefunctions valued in some commutative semiring $S$ with involution, replacing the field with involution $\complexs$ used in conventional quantum theory. In the process, the dagger compact category $\fdHilbCategory$ of finite-dimensional complex Hilbert spaces was replaced by the symmetric monoidal category $\RMatCategory{S}$ of free finite-dimensional $S$-semimodules, equipped with the dagger compact structure given by the involution of $S$ (which generalises complex conjugation). We have also postulated classical non-determinism to arise via a generalisation of the Born rule, as embodied by the CP* construction, and we have shown that our construction yields special cases of $R$-probabilistic theories, as defined in the recently introduced framework of Categorical Probabilistic Theories (here $R$ is the sub-semiring of $S$ given by the positive elements, generalising the probabilistic semiring $\reals^+$ modelling classical non-determinism in conventional quantum theory).

In subsequent Sections, we have shown our framework to be expressive enough to capture many toy models which have appeared in the literature in (more or less) recent years. Aside from real quantum theory and relational quantum theory, which have already found their special place in the heart of categorical quantum mechanicians and operational probabilistic theorists, we considered hyperbolic quantum theory, $p$-adic quantum theory and parity quantum theory (the $\integersMod{2}$ case of modal quantum theory), all interesting enough to deserve their own place in our growing zoo of categorical toy models. 

We have also introduced two new families of toy models, one based on quadratic extensions of finite field (finite-field quantum theory), and the other based on tropical semirings (tropical quantum theory). While tropical quantum theory proves to be a variant on relational quantum theory, finite-field quantum theory is of independent interest: it boasts a rich phase group which allows non-trivial quantum protocols to be implemented, while at the same time remaining fully local, as well as amenable to treatment with tools from finite geometry.

\paragraph{Future work.} This work leaves a number of directions open to future investigation. Firstly, some of the toy theories presented in this work have barely had their surface scratched from the point of view of Categorical Quantum Mechanics: an in-depth study of the categorical features they possess (e.g. unitaries, measurements/preparations, $\dagger$-Frobenius algebras, complementary and strongly complementary observables) will be a priority in further developments, together with a more thorough understanding of which quantum protocols can be implemented in them. 

Secondly, this work mainly focussed on existing toy models, or variations thereof, to show that the framework we presented truly is expressive enough for its intended purpose. However, there are many other examples of semirings, rings and fields that could potentially produce interesting and unexpected features, and we expect our zoo to continue growing in the coming years.

Finally, it was not possible, for reasons of space, to explore the applications of finite geometry to finite-field quantum theory, despite the promise of interesting connections between toy quantum systems and finite projective spaces. Similarly, it was not possible to establish whether tropical quantum theory is always local. A thorough exploration of these matters is left to future work.

\subparagraph*{Acknowledgements.}
The author would like to thank Samson Abramsky and Fabrizio Romano Genovese for comments, suggestions and useful discussions, as well as Sukrita Chatterji and Nicol\`o Chiappori for their support. Funding from EPSRC and Trinity College is gratefully acknowledged.

\newpage
\bibliographystyle{eptcs}
%\bibliography{biblio}

\begin{thebibliography}{10}
\providecommand{\bibitemdeclare}[2]{}
\providecommand{\surnamestart}{}
\providecommand{\surnameend}{}
\providecommand{\urlprefix}{Available at }
\providecommand{\url}[1]{\texttt{#1}}
\providecommand{\href}[2]{\texttt{#2}}
\providecommand{\urlalt}[2]{\href{#1}{#2}}
\providecommand{\doi}[1]{doi:\urlalt{http://dx.doi.org/#1}{#1}}
\providecommand{\bibinfo}[2]{#2}


\bibitemdeclare{article}{Abramsky2013}
\bibitem{Abramsky2013}
\bibinfo{author}{Samson \surnamestart Abramsky\surnameend}
  (\bibinfo{year}{2013}): \emph{\bibinfo{title}{{Relational Hidden Variables
  and Non-Locality}}}.
\newblock {\sl \bibinfo{journal}{Studia Logica}}
  \bibinfo{volume}{101}(\bibinfo{number}{2}), pp. \bibinfo{pages}{411--452},
  \doi{10.1007/s11225-013-9477-4}.

\bibitemdeclare{article}{Abramsky2011}
\bibitem{Abramsky2011}
\bibinfo{author}{Samson \surnamestart Abramsky\surnameend} \&
  \bibinfo{author}{Adam \surnamestart Brandenburger\surnameend}
  (\bibinfo{year}{2011}): \emph{\bibinfo{title}{{The sheaf-theoretic structure
  of non-locality and contextuality}}}.
\newblock {\sl \bibinfo{journal}{New Journal of Physics}} \bibinfo{volume}{13},
  \doi{10.1088/1367-2630/13/11/113036}.

\bibitemdeclare{incollection}{Abramsky2014}
\bibitem{Abramsky2014}
\bibinfo{author}{Samson \surnamestart Abramsky\surnameend} \&
  \bibinfo{author}{Adam \surnamestart Brandenburger\surnameend}
  (\bibinfo{year}{2014}): \emph{\bibinfo{title}{{An Operational Interpretation
  of Negative Probabilities and No-Signalling Models}}}.
\newblock In: {\sl \bibinfo{booktitle}{Horizons of the Mind. A Tribute to
  Prakash Panangaden.}}, pp. \bibinfo{pages}{59--75},
  \doi{10.1007/978-3-319-06880-0{\_}3}.

\bibitemdeclare{inproceedings}{Abramsky2004}
\bibitem{Abramsky2004}
\bibinfo{author}{S.~\surnamestart Abramsky\surnameend} \&
  \bibinfo{author}{B.~\surnamestart Coecke\surnameend} (\bibinfo{year}{2004}):
  \emph{\bibinfo{title}{{A categorical semantics of quantum protocols}}}.
\newblock In: {\sl \bibinfo{booktitle}{Proceedings of the 19th Annual IEEE
  Symposium on Logic in Computer Science, 2004.}}, \bibinfo{publisher}{IEEE},
  pp. \bibinfo{pages}{415--425}, \doi{10.1109/LICS.2004.1319636}.

\bibitemdeclare{article}{Abramsky2012c}
\bibitem{Abramsky2012c}
\bibinfo{author}{Samson \surnamestart Abramsky\surnameend} \&
  \bibinfo{author}{Lucien \surnamestart Hardy\surnameend}
  (\bibinfo{year}{2012}): \emph{\bibinfo{title}{{Logical Bell inequalities}}}.
\newblock {\sl \bibinfo{journal}{Physical Review A}}
  \bibinfo{volume}{85}(\bibinfo{number}{6}), p. \bibinfo{pages}{062114},
  \doi{10.1103/PhysRevA.85.062114}.

\bibitemdeclare{article}{Abramsky2012}
\bibitem{Abramsky2012}
\bibinfo{author}{Samson \surnamestart Abramsky\surnameend} \&
  \bibinfo{author}{Chris \surnamestart Heunen\surnameend}
  (\bibinfo{year}{2012}): \emph{\bibinfo{title}{{Operational theories and
  Categorical quantum mechanics}}}.
\newblock {\sl \bibinfo{journal}{Logic and Algebraic Structures in Quantum
  Computing}}, \doi{10.1017/CBO9781139519687.007}.

\bibitemdeclare{article}{Araki1980}
\bibitem{Araki1980}
\bibinfo{author}{Huzihiro \surnamestart Araki\surnameend}
  (\bibinfo{year}{1980}): \emph{\bibinfo{title}{{On a characterization of the
  state space of quantum mechanics}}}.
\newblock {\sl \bibinfo{journal}{Communications in Mathematical Physics}}
  \bibinfo{volume}{75}(\bibinfo{number}{1}), pp. \bibinfo{pages}{1--24},
  \doi{10.1007/BF01962588}.

\bibitemdeclare{article}{Backens2014}
\bibitem{Backens2014}
\bibinfo{author}{Miriam \surnamestart Backens\surnameend}
  (\bibinfo{year}{2014}): \emph{\bibinfo{title}{{The ZX-calculus is complete
  for stabilizer quantum mechanics}}}.
\newblock {\sl \bibinfo{journal}{New Journal of Physics}}
  \bibinfo{volume}{16}(\bibinfo{number}{9}),
  \doi{10.1088/1367-2630/16/9/093021}.

\bibitemdeclare{article}{Backens2015}
\bibitem{Backens2015}
\bibinfo{author}{Miriam \surnamestart Backens\surnameend} \&
  \bibinfo{author}{Ali~Nabi \surnamestart Duman\surnameend}
  (\bibinfo{year}{2015}): \emph{\bibinfo{title}{{A Complete Graphical Calculus
  for Spekkens' Toy Bit Theory}}}.
\newblock {\sl \bibinfo{journal}{Foundations of Physics}}
  \bibinfo{volume}{46}(\bibinfo{number}{1}), pp. \bibinfo{pages}{70--103},
  \doi{10.1007/s10701-015-9957-7}.

\bibitemdeclare{article}{Baez2012}
\bibitem{Baez2012}
\bibinfo{author}{John~C. \surnamestart Baez\surnameend} (\bibinfo{year}{2012}):
  \emph{\bibinfo{title}{{Division Algebras and Quantum Theory}}}.
\newblock {\sl \bibinfo{journal}{Foundations of Physics}}
  \bibinfo{volume}{42}(\bibinfo{number}{7}), pp. \bibinfo{pages}{819--855},
  \doi{10.1007/s10701-011-9566-z}.

\bibitemdeclare{article}{DeBeaudrap2014}
\bibitem{DeBeaudrap2014}
\bibinfo{author}{Niel\surnamestart de Beaudrap\surnameend} (\bibinfo{year}{2014}):
  \emph{\bibinfo{title}{{On computation with 'probabilities' modulo k}}}.

\bibitemdeclare{article}{Belenchia2012}
\bibitem{Belenchia2012}
\bibinfo{author}{Alessio \surnamestart Belenchia\surnameend},
  \bibinfo{author}{Giacomo~Mauro \surnamestart D'Ariano\surnameend} \&
  \bibinfo{author}{Paolo \surnamestart Perinotti\surnameend}
  (\bibinfo{year}{2012}): \emph{\bibinfo{title}{{Universality of Computation in
  Real Quantum Theory}}}.
\newblock \doi{10.1209/0295-5075/104/20006}.

\bibitemdeclare{article}{Catani2017}
\bibitem{Catani2017}
\bibinfo{author}{Lorenzo \surnamestart Catani\surnameend} \&
  \bibinfo{author}{Dan~E. \surnamestart Browne\surnameend}
  (\bibinfo{year}{2017}): \emph{\bibinfo{title}{{Spekkens' toy model in all
  dimensions and its relationship with stabilizer quantum mechanics}}}.

\bibitemdeclare{article}{Chiribella2014}
\bibitem{Chiribella2014}
\bibinfo{author}{Giulio \surnamestart Chiribella\surnameend}
  (\bibinfo{year}{2014}): \emph{\bibinfo{title}{{Dilation of states and
  processes in operational-probabilistic theories}}}.
\newblock \doi{10.4204/EPTCS.172.1}.

\bibitemdeclare{article}{Chiribella2010}
\bibitem{Chiribella2010}
\bibinfo{author}{Giulio \surnamestart Chiribella\surnameend},
  \bibinfo{author}{Giacomo~Mauro \surnamestart D'Ariano\surnameend} \&
  \bibinfo{author}{Paolo \surnamestart Perinotti\surnameend}
  (\bibinfo{year}{2010}): \emph{\bibinfo{title}{{Probabilistic theories with
  purification}}}.
\newblock {\sl \bibinfo{journal}{Physical Review A - Atomic, Molecular, and
  Optical Physics}} \bibinfo{volume}{81}(\bibinfo{number}{6}),
  \doi{10.1103/PhysRevA.81.062348}.

\bibitemdeclare{article}{Chiribella2011}
\bibitem{Chiribella2011}
\bibinfo{author}{Giulio \surnamestart Chiribella\surnameend},
  \bibinfo{author}{Giacomo~Mauro \surnamestart D'Ariano\surnameend} \&
  \bibinfo{author}{Paolo \surnamestart Perinotti\surnameend}
  (\bibinfo{year}{2011}): \emph{\bibinfo{title}{{Informational derivation of
  quantum theory}}}.
\newblock {\sl \bibinfo{journal}{Physical Review A - Atomic, Molecular, and
  Optical Physics}} \bibinfo{volume}{84}(\bibinfo{number}{1}), pp.
  \bibinfo{pages}{1--39}, \doi{10.1103/PhysRevA.84.012311}.

\bibitemdeclare{article}{Clifford1871}
\bibitem{Clifford1871}
\bibinfo{author}{\surnamestart Clifford\surnameend} (\bibinfo{year}{1871}):
  \emph{\bibinfo{title}{{Preliminary Sketch of Biquaternions}}}.
\newblock {\sl \bibinfo{journal}{Proceedings of the London Mathematical
  Society}} \bibinfo{volume}{s1-4}(\bibinfo{number}{1}), pp.
  \bibinfo{pages}{381--395}, \doi{10.1112/plms/s1-4.1.381}.

\bibitemdeclare{article}{Coecke2008a}
\bibitem{Coecke2008a}
\bibinfo{author}{Bob \surnamestart Coecke\surnameend} (\bibinfo{year}{2008}):
  \emph{\bibinfo{title}{{Axiomatic Description of Mixed States From Selinger's
  CPM-construction}}}.
\newblock {\sl \bibinfo{journal}{Electronic Notes in Theoretical Computer
  Science}} \bibinfo{volume}{210}, pp. \bibinfo{pages}{3--13},
  \doi{10.1016/j.entcs.2008.04.014}.

\bibitemdeclare{article}{Coecke2011}
\bibitem{Coecke2011}
\bibinfo{author}{Bob \surnamestart Coecke\surnameend} \& \bibinfo{author}{Ross
  \surnamestart Duncan\surnameend} (\bibinfo{year}{2011}):
  \emph{\bibinfo{title}{{Interacting quantum observables: Categorical algebra
  and diagrammatics}}}.
\newblock {\sl \bibinfo{journal}{New Journal of Physics}} \bibinfo{volume}{13},
  \doi{10.1088/1367-2630/13/4/043016}.

\bibitemdeclare{article}{Coecke2012c}
\bibitem{Coecke2012c}
\bibinfo{author}{Bob \surnamestart Coecke\surnameend}, \bibinfo{author}{Ross
  \surnamestart Duncan\surnameend}, \bibinfo{author}{Aleks \surnamestart
  Kissinger\surnameend} \& \bibinfo{author}{Quanlong \surnamestart
  Wang\surnameend} (\bibinfo{year}{2012}): \emph{\bibinfo{title}{{Strong
  complementarity and non-locality in categorical quantum mechanics}}}.
\newblock {\sl \bibinfo{journal}{Proceedings of the 2012 27th Annual ACM/IEEE
  Symposium on Logic in Computer Science, LICS 2012}}, pp.
  \bibinfo{pages}{245--254}, \doi{10.1109/LICS.2012.35}.

\bibitemdeclare{article}{Coecke2012a}
\bibitem{Coecke2012a}
\bibinfo{author}{Bob \surnamestart Coecke\surnameend} \& \bibinfo{author}{Bill
  \surnamestart Edwards\surnameend} (\bibinfo{year}{2012}):
  \emph{\bibinfo{title}{{Spekkens's toy theory as a category of processes}}}.
\newblock {\sl \bibinfo{journal}{Proceedings of Symposia in Applied
  Mathematics}} \bibinfo{volume}{71}, \doi{ http://dx.doi.org/10.1090/psapm/071/602}.

\bibitemdeclare{article}{Coecke2010b}
\bibitem{Coecke2010b}
\bibinfo{author}{Bob \surnamestart Coecke\surnameend}, \bibinfo{author}{Bill
  \surnamestart Edwards\surnameend} \& \bibinfo{author}{Robert~W. \surnamestart
  Spekkens\surnameend} (\bibinfo{year}{2010}): \emph{\bibinfo{title}{{Phase
  groups and the origin of non-locality for qubits}}}, \doi{10.1016/j.entcs.2011.01.021}.

\bibitemdeclare{article}{Coecke2012d}
\bibitem{Coecke2012d}
\bibinfo{author}{Bob \surnamestart Coecke\surnameend} \& \bibinfo{author}{Chris
  \surnamestart Heunen\surnameend} (\bibinfo{year}{2012}):
  \emph{\bibinfo{title}{{Pictures of complete positivity in arbitrary
  dimension}}}.
\newblock {\sl \bibinfo{journal}{Electronic Proceedings in Theoretical Computer
  Science}} \bibinfo{volume}{95}, pp. \bibinfo{pages}{27--35},
  \doi{10.4204/EPTCS.95.4}.

\bibitemdeclare{article}{Coecke2014a}
\bibitem{Coecke2014a}
\bibinfo{author}{Bob \surnamestart Coecke\surnameend}, \bibinfo{author}{Chris
  \surnamestart Heunen\surnameend} \& \bibinfo{author}{Aleks \surnamestart
  Kissinger\surnameend} (\bibinfo{year}{2014}):
  \emph{\bibinfo{title}{{Categories of quantum and classical channels}}}.
\newblock {\sl \bibinfo{journal}{Quantum Information Processing}}, pp.
  \bibinfo{pages}{1--31}, \doi{10.1007/s11128-014-0837-4}.

\bibitemdeclare{article}{Coecke2015}
\bibitem{Coecke2015}
\bibinfo{author}{Bob \surnamestart Coecke\surnameend} \& \bibinfo{author}{Aleks
  \surnamestart Kissinger\surnameend} (\bibinfo{year}{2015}):
  \emph{\bibinfo{title}{{Categorical Quantum Mechanics I: Causal Quantum
  Processes}}}.

\bibitemdeclare{book}{Coecke2016a}
\bibitem{Coecke2016a}
\bibinfo{author}{Bob \surnamestart Coecke\surnameend} \& \bibinfo{author}{Aleks
  \surnamestart Kissinger\surnameend} (\bibinfo{year}{2017}):
  \emph{\bibinfo{title}{{Picturing Quantum Processes}}}.
\newblock \bibinfo{publisher}{Cambridge University Press},
  \doi{10.1017/9781316219317}.

\bibitemdeclare{article}{Coecke2013b}
\bibitem{Coecke2013b}
\bibinfo{author}{Bob \surnamestart Coecke\surnameend}, \bibinfo{author}{Dusko
  \surnamestart Pavlovic\surnameend} \& \bibinfo{author}{Jamie \surnamestart
  Vicary\surnameend} (\bibinfo{year}{2013}): \emph{\bibinfo{title}{{A new
  description of orthogonal bases}}}.
\newblock {\sl \bibinfo{journal}{Mathematical Structures in Computer Science}}
  \bibinfo{volume}{23}(\bibinfo{number}{03}), \doi{10.1017/S0960129512000047}.

\bibitemdeclare{article}{Disilvestro2016}
\bibitem{Disilvestro2016}
\bibinfo{author}{Leonardo \surnamestart Disilvestro\surnameend} \&
  \bibinfo{author}{Damian \surnamestart Markham\surnameend}
  (\bibinfo{year}{2016}): \emph{\bibinfo{title}{{Quantum Protocols within
  Spekkens' Toy Model}}}.

\bibitemdeclare{article}{Evans2009}
\bibitem{Evans2009}
\bibinfo{author}{Julia \surnamestart Evans\surnameend}, \bibinfo{author}{Ross
  \surnamestart Duncan\surnameend}, \bibinfo{author}{Alex \surnamestart
  Lang\surnameend} \& \bibinfo{author}{Prakash \surnamestart
  Panangaden\surnameend} (\bibinfo{year}{2009}):
  \emph{\bibinfo{title}{{Classifying all mutually unbiased bases in Rel}}}.
\newblock \urlprefix\url{http://arxiv.org/abs/0909.4453}.

\bibitemdeclare{article}{Gogioso2015g}
\bibitem{Gogioso2015g}
\bibinfo{author}{Stefano \surnamestart Gogioso\surnameend}
  (\bibinfo{year}{2015}): \emph{\bibinfo{title}{{A Bestiary of Sets and
  Relations}}}.
\newblock {\sl \bibinfo{journal}{Electronic Proceedings in Theoretical Computer
  Science}} (\bibinfo{number}{QPL 2015}), pp. \bibinfo{pages}{208--227},
  \doi{10.4204/EPTCS.195.16}.

\bibitemdeclare{article}{Gogioso2017b}
\bibitem{Gogioso2017b}
\bibinfo{author}{Stefano \surnamestart Gogioso\surnameend} \&
  \bibinfo{author}{Aleks \surnamestart Kissinger\surnameend}
  (\bibinfo{year}{2017}): \emph{\bibinfo{title}{{Fully graphical treatment of
  the quantum algorithm for the Hidden Subgroup Problem}}}.

\bibitemdeclare{article}{Gogioso2017a}
\bibitem{Gogioso2017a}
\bibinfo{author}{Stefano \surnamestart Gogioso\surnameend} \&
  \bibinfo{author}{Carlo~Maria \surnamestart Scandolo\surnameend}
  (\bibinfo{year}{2017}): \emph{\bibinfo{title}{{Categorical Probabilistic
  Theories}}}.

\bibitemdeclare{article}{Gogioso2015d}
\bibitem{Gogioso2015d}
\bibinfo{author}{Stefano \surnamestart Gogioso\surnameend} \&
  \bibinfo{author}{William \surnamestart Zeng\surnameend}
  (\bibinfo{year}{2015}): \emph{\bibinfo{title}{{Fourier transforms from
  strongly complementary observables}}}.

\bibitemdeclare{article}{Gogioso2015}
\bibitem{Gogioso2015}
\bibinfo{author}{Stefano \surnamestart Gogioso\surnameend} \&
  \bibinfo{author}{William \surnamestart Zeng\surnameend}
  (\bibinfo{year}{2015}): \emph{\bibinfo{title}{{Mermin Non-Locality in
  Abstract Process Theories}}}.
\newblock {\sl \bibinfo{journal}{Electronic Proceedings in Theoretical Computer
  Science}} (\bibinfo{number}{QPL 2015}), pp. \bibinfo{pages}{228--246},
  \doi{10.4204/EPTCS.195.17}.

\bibitemdeclare{article}{Gogioso2017c}
\bibitem{Gogioso2017c}
\bibinfo{author}{Stefano \surnamestart Gogioso\surnameend} \&
  \bibinfo{author}{William \surnamestart Zeng\surnameend}
  (\bibinfo{year}{2017}): \emph{\bibinfo{title}{{Generalised Mermin-type
  non-locality arguments}}}.

\bibitemdeclare{incollection}{HereBeDragons}
\bibitem{HereBeDragons}
\bibinfo{author}{Andrew J. \surnamestart Hamilton\surnameend, Robert M. \surnamestart May\surnameend \& Edward K. \surnamestart Waters\surnameend} (\bibinfo{year}{2015}):
  \emph{\bibinfo{title}{{Zoology: Here be dragons}}}. 
\newblock {\sl \bibinfo{journal}{Nature}} (\bibinfo{number}{520}), pp. \bibinfo{pages}{42--43},
  \doi{10.1038/520042a}.

\bibitemdeclare{incollection}{Hardy2009}
\bibitem{Hardy2009}
\bibinfo{author}{Lucien \surnamestart Hardy\surnameend} (\bibinfo{year}{2009}):
  \emph{\bibinfo{title}{{Foliable Operational Structures for General
  Probabilistic Theories}}}. 
\newblock In \bibinfo{editor}{Hans \surnamestart Halvorson\surnameend},
  editor: {\sl \bibinfo{booktitle}{Deep Beauty}}, \bibinfo{publisher}{Cambridge
  University Press}, \doi{10.1017/CBO9780511976971.013}.

\bibitemdeclare{article}{Hardy2016}
\bibitem{Hardy2016}
\bibinfo{author}{Lucien \surnamestart Hardy\surnameend} (\bibinfo{year}{2016}):
  \emph{\bibinfo{title}{{Reconstructing Quantum Theory}}}.
\newblock pp. \bibinfo{pages}{223--248}, \doi{10.1007/978-94-017-7303-4{\_}7}.

\bibitemdeclare{article}{Heunen2015}
\bibitem{Heunen2015}
\bibinfo{author}{Chris \surnamestart Heunen\surnameend} \&
  \bibinfo{author}{Sean \surnamestart Tull\surnameend} (\bibinfo{year}{2015}):
  \emph{\bibinfo{title}{{Categories of relations as models of quantum
  theory}}}.
\newblock {\sl \bibinfo{journal}{Electronic Proceedings in Theoretical Computer
  Science}} (\bibinfo{number}{Qpl2015}),
  \doi{10.4204/EPTCS.195.18}.

\bibitemdeclare{book}{Heunen2016a}
\bibitem{Heunen2016a}
\bibinfo{author}{Chris \surnamestart Heunen\surnameend} \&
  \bibinfo{author}{Jamie \surnamestart Vicary\surnameend}
  (\bibinfo{year}{2017}): \emph{\bibinfo{title}{{Introduction to Categorical
  Quantum Mechanics}}}.
\newblock \bibinfo{publisher}{OUP}.

\bibitemdeclare{book}{Hilbert1998}
\bibitem{Hilbert1998}
\bibinfo{author}{David \surnamestart Hilbert\surnameend}
  (\bibinfo{year}{1998}): \emph{\bibinfo{title}{{The Theory of Algebraic Number
  Fields}}}.
\newblock \bibinfo{publisher}{Springer Berlin Heidelberg},
  \bibinfo{address}{Berlin, Heidelberg}, \doi{10.1007/978-3-662-03545-0}.

\bibitemdeclare{article}{Jordan1934}
\bibitem{Jordan1934}
\bibinfo{author}{P.~\surnamestart Jordan\surnameend},
  \bibinfo{author}{J.~\surnamestart v.~Neumann\surnameend} \&
  \bibinfo{author}{E.~\surnamestart Wigner\surnameend} (\bibinfo{year}{1934}):
  \emph{\bibinfo{title}{{On an Algebraic Generalization of the Quantum
  Mechanical Formalism}}}.
\newblock {\sl \bibinfo{journal}{The Annals of Mathematics}}
  \bibinfo{volume}{35}(\bibinfo{number}{1}),
  \doi{10.2307/1968117}.

\bibitemdeclare{article}{Khrennikov1991}
\bibitem{Khrennikov1991}
\bibinfo{author}{A.~Yu. \surnamestart Khrennikov\surnameend}
  (\bibinfo{year}{1991}): \emph{\bibinfo{title}{p ‐adic quantum mechanics
  with p ‐adic valued functions}}.
\newblock {\sl \bibinfo{journal}{Journal of Mathematical Physics}}
  \bibinfo{volume}{32}(\bibinfo{number}{4}), pp. \bibinfo{pages}{932--937},
  \doi{10.1063/1.529353}.

\bibitemdeclare{article}{Khrennikov1993}
\bibitem{Khrennikov1993}
\bibinfo{author}{A.~Yu. \surnamestart Khrennikov\surnameend}
  (\bibinfo{year}{1993}): \emph{\bibinfo{title}{{p-Adic probability theory and
  its applications. The principle of statistical stabilization of
  frequencies}}}.
\newblock {\sl \bibinfo{journal}{Theoretical and Mathematical Physics}}
  \bibinfo{volume}{97}(\bibinfo{number}{3}), pp. \bibinfo{pages}{1340--1348},
  \doi{10.1007/BF01015763}.

\bibitemdeclare{article}{Khrennikov2003}
\bibitem{Khrennikov2003}
\bibinfo{author}{Andrei \surnamestart Khrennikov\surnameend}
  (\bibinfo{year}{2003}): \emph{\bibinfo{title}{{Hyperbolic quantum
  mechanics}}}.
\newblock {\sl \bibinfo{journal}{Advances in Applied Clifford Algebras}}
  \bibinfo{volume}{13}(\bibinfo{number}{1}), pp. \bibinfo{pages}{1--9},
  \doi{10.1007/s00006-003-0001-1}.

\bibitemdeclare{article}{Khrennikov2010}
\bibitem{Khrennikov2010}
\bibinfo{author}{Andrei \surnamestart Khrennikov\surnameend}
  (\bibinfo{year}{2010}): \emph{\bibinfo{title}{{Representation of
  Probabilistic Data by Quantum-Like Hyperbolic Amplitudes}}}.
\newblock {\sl \bibinfo{journal}{Advances in Applied Clifford Algebras}}
  \bibinfo{volume}{20}(\bibinfo{number}{1}), pp. \bibinfo{pages}{43--56},
  \doi{10.1007/s00006-008-0139-y}.

\bibitemdeclare{article}{Marsden}
\bibitem{Marsden}
\bibinfo{author}{Daniel \surnamestart Marsden\surnameend}:
  \emph{\bibinfo{title}{{A Graph Theoretic Perspective on CPM(Rel)}}}.
\newblock {\sl \bibinfo{journal}{Electronic Proceedings in Theoretical Computer
  Science}} (\bibinfo{number}{QPL 2015}), \doi{10.4204/EPTCS.195.20}.

\bibitemdeclare{article}{Maslov1987}
\bibitem{Maslov1987}
\bibinfo{author}{V~P \surnamestart Maslov\surnameend} (\bibinfo{year}{1987}):
  \emph{\bibinfo{title}{{On a new principle of superposition for optimization
  problems}}}.
\newblock {\sl \bibinfo{journal}{Russian Mathematical Surveys}}
  \bibinfo{volume}{42}(\bibinfo{number}{3}), pp. \bibinfo{pages}{43--54},
  \doi{10.1070/RM1987v042n03ABEH001439}.

\bibitemdeclare{article}{Nyman2011}
\bibitem{Nyman2011}
\bibinfo{author}{Peter \surnamestart Nyman\surnameend} (\bibinfo{year}{2011}):
  \emph{\bibinfo{title}{{On the Consistency of the Quantum-Like Representation
  Algorithm for Hyperbolic Interference}}}.
\newblock {\sl \bibinfo{journal}{Advances in Applied Clifford Algebras}}
  \bibinfo{volume}{21}(\bibinfo{number}{4}), pp. \bibinfo{pages}{799--811},
  \doi{10.1007/s00006-011-0287-3}.

\bibitemdeclare{article}{Palmer2016}
\bibitem{Palmer2016}
\bibinfo{author}{T.~N. \surnamestart Palmer\surnameend} (\bibinfo{year}{2016}):
  \emph{\bibinfo{title}{{Invariant Set Theory}}}.
\newblock \urlprefix\url{http://arxiv.org/abs/1605.01051}.

\bibitemdeclare{article}{Palmer2016a}
\bibitem{Palmer2016a}
\bibinfo{author}{T.~N. \surnamestart Palmer\surnameend} (\bibinfo{year}{2016}):
  \emph{\bibinfo{title}{{{\$}p{\$}-adic Distance, Finite Precision and Emergent
  Superdeterminism: A Number-Theoretic Consistent-Histories Approach to Local
  Quantum Realism}}}.

\bibitemdeclare{incollection}{Pin1998}
\bibitem{Pin1998}
\bibinfo{author}{Jean-Eric \surnamestart Pin\surnameend}
  (\bibinfo{year}{1998}): \emph{\bibinfo{title}{{Tropical semirings}}}.
\newblock In \bibinfo{editor}{Jeremy \surnamestart Gunawardena\surnameend},
  editor: {\sl \bibinfo{booktitle}{Idempotency}}, \bibinfo{publisher}{Cambridge
  University Press}, \bibinfo{address}{Cambridge}, pp. \bibinfo{pages}{50--69},
  \doi{10.1017/CBO9780511662508.004}.

\bibitemdeclare{article}{Pavlovic2009}
\bibitem{Pavlovic2009}
\bibinfo{author}{Dusko \surnamestart Pavlovic\surnameend}
  (\bibinfo{year}{2009}): \emph{\bibinfo{title}{{Quantum and classical
  structures in nondeterminstic computation}}}.
\newblock {\sl \bibinfo{journal}{Lecture Notes in Computer Science (including
  subseries Lecture Notes in Artificial Intelligence and Lecture Notes in
  Bioinformatics)}} \bibinfo{volume}{5494}, pp. \bibinfo{pages}{143--157},
  \doi{10.1007/978-3-642-00834-4{\_}13}.

\bibitemdeclare{article}{Rovelli1996}
\bibitem{Rovelli1996}
\bibinfo{author}{Carlo \surnamestart Rovelli\surnameend}
  (\bibinfo{year}{1996}): \emph{\bibinfo{title}{{Relational quantum
  mechanics}}}.
\newblock {\sl \bibinfo{journal}{International Journal of Theoretical Physics}}
  \bibinfo{volume}{35}(\bibinfo{number}{8}), pp. \bibinfo{pages}{1637--1678},
  \doi{10.1007/BF02302261}.

\bibitemdeclare{article}{Ruelle1989}
\bibitem{Ruelle1989}
\bibinfo{author}{Ph. \surnamestart Ruelle\surnameend},
  \bibinfo{author}{E.~\surnamestart Thiran\surnameend},
  \bibinfo{author}{D.~\surnamestart Verstegen\surnameend} \&
  \bibinfo{author}{J.~\surnamestart Weyers\surnameend} (\bibinfo{year}{1989}):
  \emph{\bibinfo{title}{{Quantum mechanics on p ‐adic fields}}}.
\newblock {\sl \bibinfo{journal}{Journal of Mathematical Physics}}
  \bibinfo{volume}{30}(\bibinfo{number}{12}), pp. \bibinfo{pages}{2854--2874},
  \doi{10.1063/1.528468}.

\bibitemdeclare{book}{FantasticBeastsAndWhereToFindThem}
\bibitem{FantasticBeastsAndWhereToFindThem}
\bibinfo{author}{Newt \surnamestart Scamander\surnameend} (\bibinfo{year}{1927}):
  \emph{\bibinfo{title}{{Fantastic Beasts and Where to Find Them}}}. 
\newblock \bibinfo{publisher}{Obscurus Books}.

\bibitemdeclare{article}{Schumacher2012}
\bibitem{Schumacher2012}
\bibinfo{author}{Benjamin \surnamestart Schumacher\surnameend} \&
  \bibinfo{author}{Michael~D. \surnamestart Westmoreland\surnameend}
  (\bibinfo{year}{2012}): \emph{\bibinfo{title}{{Modal Quantum Theory}}}.
\newblock {\sl \bibinfo{journal}{Foundations of Physics}}
  \bibinfo{volume}{42}(\bibinfo{number}{7}), pp. \bibinfo{pages}{918--925},
  \doi{10.1007/s10701-012-9650-z}.

\bibitemdeclare{incollection}{Schumacher2016}
\bibitem{Schumacher2016}
\bibinfo{author}{Benjamin \surnamestart Schumacher\surnameend} \&
  \bibinfo{author}{Michael~D. \surnamestart Westmoreland\surnameend}
  (\bibinfo{year}{2016}): \emph{\bibinfo{title}{{Almost Quantum Theory}}}.
\newblock pp. \bibinfo{pages}{45--81}, \doi{10.1007/978-94-017-7303-4{\_}3}.

\bibitemdeclare{article}{Selinger2007}
\bibitem{Selinger2007}
\bibinfo{author}{Peter \surnamestart Selinger\surnameend}
  (\bibinfo{year}{2007}): \emph{\bibinfo{title}{{Dagger Compact Closed
  Categories and Completely Positive Maps}}}.
\newblock {\sl \bibinfo{journal}{Electronic Notes in Theoretical Computer
  Science}} \bibinfo{volume}{170}, pp. \bibinfo{pages}{139--163},
  \doi{10.1016/j.entcs.2006.12.018}.

\bibitemdeclare{article}{Selinger2008}
\bibitem{Selinger2008}
\bibinfo{author}{Peter \surnamestart Selinger\surnameend}
  (\bibinfo{year}{2008}): \emph{\bibinfo{title}{{Idempotents in dagger
  categories ( extended abstract )}}}.
\newblock {\sl \bibinfo{journal}{Electronic Notes in Theoretical Computer
  Science}} \bibinfo{volume}{210}, pp. \bibinfo{pages}{107--122},
  \doi{10.1016/j.entcs.2008.04.021}.

\bibitemdeclare{incollection}{Simon1988}
\bibitem{Simon1988}
\bibinfo{author}{Imre \surnamestart Simon\surnameend} (\bibinfo{year}{1988}):
  \emph{\bibinfo{title}{{Recognizable sets with multiplicities in the tropical
  semiring}}}.
\newblock In: {\sl \bibinfo{booktitle}{Mathematical Foundations of Computer
  Science}}, \bibinfo{publisher}{Springer-Verlag},
  \bibinfo{address}{Berlin/Heidelberg}, pp. \bibinfo{pages}{107--120},
  \doi{10.1007/BFb0017135}.

\bibitemdeclare{article}{Simon1994}
\bibitem{Simon1994}
\bibinfo{author}{Imre \surnamestart Simon\surnameend} (\bibinfo{year}{1994}):
  \emph{\bibinfo{title}{{On semigroups of matrices over the tropical
  semiring}}}.
\newblock {\sl \bibinfo{journal}{RAIRO - Theoretical Informatics and
  Applications}} \bibinfo{volume}{28}(\bibinfo{number}{3-4}), pp.
  \bibinfo{pages}{277--294}, \doi{10.1051/ita/1994283-402771}.

\bibitemdeclare{article}{Spekkens2007}
\bibitem{Spekkens2007}
\bibinfo{author}{Robert~W. \surnamestart Spekkens\surnameend}
  (\bibinfo{year}{2007}): \emph{\bibinfo{title}{{Evidence for the epistemic
  view of quantum states: A toy theory}}}.
\newblock {\sl \bibinfo{journal}{Physical Review A}}
  \bibinfo{volume}{75}(\bibinfo{number}{3}), p. \bibinfo{pages}{032110},
  \doi{10.1103/PhysRevA.75.032110}.

\bibitemdeclare{article}{Speyer2009}
\bibitem{Speyer2009}
\bibinfo{author}{David \surnamestart Speyer\surnameend} \&
  \bibinfo{author}{Bernd \surnamestart Sturmfels\surnameend}
  (\bibinfo{year}{2009}): \emph{\bibinfo{title}{{Tropical Mathematics}}}.
\newblock {\sl \bibinfo{journal}{Mathematics Magazine}}
  \bibinfo{volume}{82}(\bibinfo{number}{3}), pp. \bibinfo{pages}{163--173},
  \doi{10.4169/193009809X468760}.

\bibitemdeclare{inproceedings}{Vicary2012a}
\bibitem{Vicary2012a}
\bibinfo{author}{Jamie \surnamestart Vicary\surnameend} (\bibinfo{year}{2012}):
  \emph{\bibinfo{title}{{Topological structure of quantum algorithms}}}.
\newblock In: {\sl \bibinfo{booktitle}{Proceedings of the 2013 28th Annual
  ACM/IEEE Symposium on Logic in Computer Science}},
  \doi{10.1109/LICS.2013.14}.

\bibitemdeclare{article}{Vladimirov1989}
\bibitem{Vladimirov1989}
\bibinfo{author}{V.~S. \surnamestart Vladimirov\surnameend} \&
  \bibinfo{author}{I.~V. \surnamestart Volovich\surnameend}
  (\bibinfo{year}{1989}): \emph{\bibinfo{title}{p-adic quantum mechanics}}.
\newblock {\sl \bibinfo{journal}{Communications in Mathematical Physics}}
  \bibinfo{volume}{123}(\bibinfo{number}{4}), pp. \bibinfo{pages}{659--676},
  \doi{10.1007/BF01218590}.

\bibitemdeclare{article}{Wilce2016}
\bibitem{Wilce2016}
\bibinfo{author}{Alexander \surnamestart Wilce\surnameend}
  (\bibinfo{year}{2016}): \emph{\bibinfo{title}{{A Royal Road to Quantum Theory
  (or Thereabouts)}}}.
\newblock \doi{10.4204/EPTCS.236.16}.

\bibitemdeclare{article}{Wootters1990}
\bibitem{Wootters1990}
\bibinfo{author}{William~K \surnamestart Wootters\surnameend}
  (\bibinfo{year}{1990}): \emph{\bibinfo{title}{{Local accessibility of quantum
  states}}}.
\newblock {\sl \bibinfo{journal}{Complexity, entropy and the physics of
  information}} \bibinfo{volume}{8}, pp. \bibinfo{pages}{39--46}.

\bibitemdeclare{article}{Zeng2015}
\bibitem{Zeng2015}
\bibinfo{author}{William \surnamestart Zeng\surnameend} (\bibinfo{year}{2015}):
  \emph{\bibinfo{title}{{Models of Quantum Algorithms in Sets and Relations}}}.

\end{thebibliography}

\newpage
\appendix

\section{Proofs}

\paragraph{Proof of Theorem \ref{thm_localityFields}.}
%\begin{proof}
Theorem 5.4 from Ref. \cite{Abramsky2011} states that all no-signalling empirical models over the field $\reals$ admit a local hidden variable model in terms of signed probabilities. Although the original result was proven for $\reals$, close inspection reveals that it holds for no-signalling empirical models over any field $k$: as a consequence, Bell-type measurement scenarios in $R$-probabilistic theories where $R$ is a field give rise to no-signalling empirical models admitting local hidden variable models. Finally, $R$-probabilistic theories have a sub-SMC of finite $R$-probabilistic classical systems, with all $R$-distributions as normalised states and all $R$-stochastic maps as normalised processes: as a consequence, all local hidden variable models valued in $R$ can be realised in any and all $R$-probabilistic theories. \qed
%\end{proof}

\paragraph{Proof of Theorem \ref{thm_RprobabilisticCPStarCategories}.}
%\begin{proof}
In order for $\CPStarCategory{\RMatCategory{S}}$ to be $R$-probabilistic under the $\CMonCategory$-enrichment of $\RMatCategory{S}$, we need to show that it satisfies the following three conditions:
\begin{enumerate}
	\item[(i)] there is a full sub-SMC $\classicalSubcategory{\CPStarCategory{\RMatCategory{S}}}$ of $\CPStarCategory{\RMatCategory{S}}$ which is equivalent to $\RMatCategory{R}$;
	\item[(ii)] the $\CMonCategory$-enrichment of $\RMatCategory{S}$ must restrict to a well-defined $\CMonCategory$-enrichment for $\CPStarCategory{\RMatCategory{S}}$, which coincides on $\classicalSubcategory{\CPStarCategory{\RMatCategory{S}}}$ with the enrichment of $\RMatCategory{R}$;
	\item[(iii)] the SMC $\CPStarCategory{\RMatCategory{S}}$ comes with an environment structure which restricts to the the canonical one from $\RMatCategory{R}$ on the full subcategory $\classicalSubcategory{\CPStarCategory{\RMatCategory{S}}}$.
\end{enumerate}

\noindent Firstly, we show that the $\CMonCategory$-enrichment of $\RMatCategory{S}$ restricts to a well-defined $\CMonCategory$-enrichment for $\CPStarCategory{\RMatCategory{S}}$. Because $\RMatCategory{S}$ is a category of matrices, this is in turn true if and only if the scalars of $\CPStarCategory{\RMatCategory{S}}$ are closed under addition in $\RMatCategory{S}$, i.e. if and only if they form a sub-semiring of $S$ (they are always necessarily closed under multiplication). To see that this is true, it suffices to show that the scalars of $\CPStarCategory{\RMatCategory{S}}$ form exactly the sub-semiring $R$ of positive elements of $S$ (we have to show it anyway, if we want our theory to be $R$-probabilistic!). Indeed, the generic scalar of $\CPStarCategory{\RMatCategory{S}}$ takes the form $\trace{\,S^D} \circ \CPMdoubled{\ket{\psi}} = \sum_{d=1}^D p_d^\ast p_d$ for a generic state $\ket{\psi} := \sum_{d=1}^D  p_d \ket{d}$ of $\RMatCategory{S}$.

For condition (i), consider the full-subcategory $\classicalSubcategory{\CPStarCategory{\RMatCategory{S}}}$ of $\CPStarCategory{\RMatCategory{S}}$ spanned by those objects in the form $(S^X,\decoh{\hbox{\begin{tikzpicture} [scale=1.2,transform shape] %% DO NOT CHANGE

\def\deltax{0.3} %% CAN BE CHANGED
\def\deltay{0.5} %% DO NOT CHANGE

%\path[use as bounding box] (-\deltax,-\deltay) rectangle (\deltax,\deltay);

\node [dot, fill=\Zbwcolour] (mult) at (0,0) {};
%\node (mult_label_out) at (0,+\deltay) {};
%\draw[-] (mult) to (mult_label_out);

%\draw (current bounding box.south west) rectangle (current bounding box.north east);
\end{tikzpicture}}\!\!_X})$, where $X$ is a finite set, $\hbox{\begin{tikzpicture} [scale=1.2,transform shape] %% DO NOT CHANGE

\def\deltax{0.3} %% CAN BE CHANGED
\def\deltay{0.5} %% DO NOT CHANGE

%\path[use as bounding box] (-\deltax,-\deltay) rectangle (\deltax,\deltay);

\node [dot, fill=\Zbwcolour] (mult) at (0,0) {};
%\node (mult_label_out) at (0,+\deltay) {};
%\draw[-] (mult) to (mult_label_out);

%\draw (current bounding box.south west) rectangle (current bounding box.north east);
\end{tikzpicture}}\!\!_X$ is the special commutative $\dagger$-Frobenius algebra on $S^X$ associated with the orthonormal basis $\ket{x}_{x \in X}$, and $\decoh{\hbox{\begin{tikzpicture} [scale=1.2,transform shape] %% DO NOT CHANGE

\def\deltax{0.3} %% CAN BE CHANGED
\def\deltay{0.5} %% DO NOT CHANGE

%\path[use as bounding box] (-\deltax,-\deltay) rectangle (\deltax,\deltay);

\node [dot, fill=\Zbwcolour] (mult) at (0,0) {};
%\node (mult_label_out) at (0,+\deltay) {};
%\draw[-] (mult) to (mult_label_out);

%\draw (current bounding box.south west) rectangle (current bounding box.north east);
\end{tikzpicture}}\!\!_X}: S^X \rightarrow S^X$ is the decoherence map for $\hbox{\begin{tikzpicture} [scale=1.2,transform shape] %% DO NOT CHANGE

\def\deltax{0.3} %% CAN BE CHANGED
\def\deltay{0.5} %% DO NOT CHANGE

%\path[use as bounding box] (-\deltax,-\deltay) rectangle (\deltax,\deltay);

\node [dot, fill=\Zbwcolour] (mult) at (0,0) {};
%\node (mult_label_out) at (0,+\deltay) {};
%\draw[-] (mult) to (mult_label_out);

%\draw (current bounding box.south west) rectangle (current bounding box.north east);
\end{tikzpicture}}\!\!_X$, which is a self-adjoint idempotent normalised CP map. Morphisms $(S^X,\decoh{\hbox{\begin{tikzpicture} [scale=1.2,transform shape] %% DO NOT CHANGE

\def\deltax{0.3} %% CAN BE CHANGED
\def\deltay{0.5} %% DO NOT CHANGE

%\path[use as bounding box] (-\deltax,-\deltay) rectangle (\deltax,\deltay);

\node [dot, fill=\Zbwcolour] (mult) at (0,0) {};
%\node (mult_label_out) at (0,+\deltay) {};
%\draw[-] (mult) to (mult_label_out);

%\draw (current bounding box.south west) rectangle (current bounding box.north east);
\end{tikzpicture}}\!\!_X}) \rightarrow (S^Y,\decoh{\hbox{\begin{tikzpicture} [scale=1.2,transform shape] %% DO NOT CHANGE

\def\deltax{0.3} %% CAN BE CHANGED
\def\deltay{0.5} %% DO NOT CHANGE

%\path[use as bounding box] (-\deltax,-\deltay) rectangle (\deltax,\deltay);

\node [dot, fill=\Zbwcolour] (mult) at (0,0) {};
%\node (mult_label_out) at (0,+\deltay) {};
%\draw[-] (mult) to (mult_label_out);

%\draw (current bounding box.south west) rectangle (current bounding box.north east);
\end{tikzpicture}}\!\!_Y})$ are exactly those in the following form, where $(f_{xy})_{x \in X, y \in Y}$ is an arbitrary matrix of scalars (i.e. elements of $R$):
\begin{equation}
\sum_{y \in Y}\sum_{x \in X} \CPMdoubled{\ket{y}}\; f_{xy}\; \CPMdoubled{\bra{x}}
\end{equation}
As a consequence, $\classicalSubcategory{\CPStarCategory{\RMatCategory{S}}}$ is equivalent to $\RMatCategory{R}$, and condition (ii) is satisfied as well.

For condition (iii), it suffices to consider the canonical environment structure given by the CP* construction. Because decoherence maps are normalised, this environment structure restricts to the canonical one of $\RMatCategory{S}$ on the full subcategory $\classicalSubcategory{\CPStarCategory{\RMatCategory{S}}}$. \qed
%\end{proof}

\paragraph{Proof of Corollary \ref{cor_phasesFFQT}.}
%\begin{proof}
We have a quadratic extension $\finiteField{p^{n}}(\sqrt{\epsilon})/\finiteField{p^n}$, with 2-element Galois group generated by the involution $\sigma~:=~\xi~\mapsto~\xi^\ast$, and corresponding field norm $N_{\finiteField{p^{n}}(\sqrt{\epsilon}) / \finiteField{p^n}}(\xi) := \xi^\ast \xi$. By Hilbert's~Theorem~90, the multiplicative group of those $\xi \in \finiteField{p^{2n}}$ such that $\xi^\ast \xi = 1$ is isomorphic to the quotient group $\finiteField{p^{n}}(\sqrt{\epsilon})^\times/\finiteField{p^n}^\times$. But $\finiteField{p^{n}}(\sqrt{\epsilon})^\times \isom \finiteField{p^{2n}}^\times$ is cyclic with $p^{2n}-1$ elements, and $\finiteField{p^n}^\times$ has $p^n-1$ elements: hence the quotient is cyclic with $(p^{2n}-1)/(p^n-1) = p^n+1$ elements, i.e. it is $\integersMod{p^n+1}$. \qed
%\end{proof}

\paragraph{Proof of Lemma \ref{lem_purescalarsFFQT}.}
%\begin{proof}
Because $\finiteField{p^{n}}(\sqrt{\epsilon})$ is a field, we have that $a^\ast a = b^\ast b$ if and only if $a = \xi b$ for some $\xi$ such that $\xi^\ast \xi = 1$, i.e. for some phase $\xi$. Equality up to phase is an equivalence relation on elements of $\finiteField{p^{n}}(\sqrt{\epsilon})$ (because phases form a group under multiplication), and there are exactly $p^n+1$ phases by Corollary \ref{cor_phasesFFQT}: as a consequence, there are exactly $(p^{2n}-1)/(p^n+1)= p^n-1$ non-zero pure scalars in $\CPStarCategory{\RMatCategory{\finiteField{p^{n}}(\sqrt{\epsilon})}}$, i.e. all the scalars are in fact pure. \qed
%\end{proof}

\paragraph{Proof of Lemma \ref{lem_HSPinFFQT}}
%\begin{proof}
By Corollary \ref{cor_phasesFFQT}, the phases of $\finiteField{p^n}(\sqrt{\epsilon})$ form the finite cyclic group $\integersMod{p^n+1}$, and hence the $\finiteField{p^n}(\sqrt{\epsilon})$-valued unitary multiplicative characters of $G$ are exactly the group homomorphisms $G \rightarrow \integersMod{p^n+1}$. The unitary multiplicative characters of a product $\prod_{k=1}^{K} \integersMod{p_k^{e_k}}$ (where $p_1,...,p_K$ are pairwise distinct primes) take the form $(g_1,...,g_K) \mapsto \goodchi_1(g_1) \cdot... \cdot \goodchi_K(g_K)$, where $(\goodchi_1,...,\goodchi_K)$ are all possible $K$-tuples where each $\goodchi_k$ is a unitary multiplicative character of the corresponding factor $\integersMod{p_k^{e_k}}$. Hence $G \isom \prod_{k=1}^{K} \integersMod{p_k^{e_k}}$ has enough multiplicative characters if and only if each factor $\integersMod{p_k^{e_k}}$ does, and in turn this is true if and only if $p_k^{e_k} | p^n+1$ for all $k=1,...,K$. The final statement about the Hidden Subgroup Problem is a consequence of the main result from Ref. \cite{Gogioso2017b}: because all positive elements are pure scalars, is is always true that $|G| = z_G^\ast z_G$ for some $z_G \in \finiteField{p^n}(\sqrt{\epsilon})$, and furthermore $|G|$ is always invertible because we must necessarily have that $p$ does not divide $|G|$ (otherwise we would get $p | p^n+1$, which is absurd).  \qed
%\end{proof}

\paragraph{Proof of Lemma \ref{lem_merminFFQT}.}
%\begin{proof}
If $q^2 | p^n+1$, we can consider the following argument. We take the subgroup of classical states to be $K \isom \integersMod{q}$, seen as the subgroup $ K = \langle (\frac{p^n+1}{q},2\frac{p^n+1}{q},...,(q-1)\frac{p^n+1}{q}) \rangle \triangleleft \integersMod{p^n+1}^{q-1}$, and we use the equation $q y = (\frac{p^n+1}{q},2\frac{p^n+1}{q},...,(q-1)\frac{p^n+1}{q})$. The equation cannot have any solution in $K$, where $q y = (0,0,...,0)$ for all $y$, but has solution $y=(\frac{p^n+1}{q^2},2\frac{p^n+1}{q^2},...,(q-1)\frac{p^n+1}{q^2})$ in the group of phase gates $\integersMod{p^n+1}^{q-1}$.
Conversely, if $p^n+1 = \prod_{k=1}^K p_k$ for distinct primes $p_1,...,p_K$, then for any classical subgroup $K$ the group of phase gates decomposes as $K \times K'$ for some $K'$, and a result of \cite{Gogioso2015} shows that no non-trivial generalised Mermin-type argument can be formulated. We have used the fact that $|\integersMod{q}|$ is always in the form $|\integersMod{q}| = q = z_q^\ast z_q$ for some $z_q \in \finiteField{p^n}(\sqrt{\epsilon})$: this is because all positive elements are pure scalars and $|\integersMod{q}|$ must be invertible ($p$ cannot divide $q$, otherwise we would get $p | p^n+1$). \qed
%\end{proof}

\paragraph{Proof of Lemma \ref{lem_tropicalSemiringPositiveEls}.}
%\begin{proof}
Let $^\ast$ be an involution for the tropical semiring $S$: $x \leq y$ implies that $x = x+y$, so that $x^\ast = x^\ast+y^\ast$ and $x^\ast \leq y^\ast$. But then $x \leq x^\ast$ implies $x^\ast \leq (x^\ast)^\ast = x$ (and similarly for $x^\ast \leq x$), so that $x^\ast = x$ is the trivial involution. Now consider the tropical semiring with trivial involution, so that the positive elements are exactly those in the form $x^2$ for some $x \in S$. But in a tropical semiring we have that $x^2+y^2 = (x+y)^2$ (as Speyer and Sturmfels put it, \inlineQuote{the Freshman's dream holds in tropical arithmetic.} \cite{Speyer2009}): hence the squares are closed under addition $+$, and form a sub-semiring. \qed
%\end{proof}

\end{document}